\documentclass[a4paper,11pt]{article}
\pdfoutput=1
\usepackage{amsmath,amsfonts,amssymb,times,mathptmx,graphicx,amsthm}
\usepackage{placeins}
\usepackage[T1]{fontenc}
\usepackage[latin1]{inputenc}
\title{\bf Sub-optimal Approaches to Heteroscedasticity in
Silicon Strip Detectors: the Lucky Model and  the  Super-Lucky Model.
 }
\author{Gregorio Landi$^a$\thanks{Corresponding
author. Gregorio.Landi}~,   Giovanni E. Landi$^b$\\
\\
\llap{$^a$} Dipartimento di Fisica e Astronomia,
Universita' di Firenze\\
Largo E. Fermi 2 (Arcetri) 50125, Firenze, Italy\\
\\
\llap{$^b$} ArchonVR S.a.g.l.,\\
Via Cisieri 3,
6900 Lugano, Switzerland.\\ \\
{ May 23, 2022}}
\date{ }
\begin{document}
\maketitle 
\begin{abstract}
The approach to heteroscedasticity of ref.~\cite{landi15}
contains a sketchy application of a sub-optimal method
of very easy implementation: the lucky model. The
supporting proof of this method could not be inserted
in~\cite{landi15}. The proof requires the analytical
forms of the probability of ref.~\cite{landi10} for
the two strip center of gravity.
However, those analytical forms
suggest also a completion of the lucky-model for
the absence of a scaling constant, relevant for
combinations of different detector types.
The advanced lucky-model (the super-lucky model)
can be directly used for track fitting in
trackers composed of non-identical
detectors. The construction of the weights for the fits
is very simple. Simulations
of track fitting with this upgraded tool show
resolution improvements also for combination of two
types of very different detectors, near to the resolutions
of the schematic model of reference~\cite{landi15}.
\end{abstract}


\newpage


\pagenumbering{arabic} \oddsidemargin 0cm  \evensidemargin 0cm


\section{Introduction}

The heteroscedasticity of silicon strip detectors requires
a detailed knowledge of the variances of the observations
(hits) to insert in weighted least squares to obtain  optimal fits~\cite{landi08,landi09}.
The signals released by a minimum ionizing particle (MIP), if
carefully analyzed, allow to extract sub-optimal effective standard
deviations defined in ref.~\cite{landi15} as lucky-model. This model
was illustrated without a detailed discussion.
A more accurate discussion of the lucky model, promised in ref.~\cite{landi15},
will be developed here. A brief recall to the
equations of ref.~\cite{landi10} is essential for that approximate
demonstration. Elements of those equations
give a suggestion for the extension of the lucky-model.
The lucky model was suggested by a strong similarity of
the parameter distributions of the schematic model~\cite{landi05}
with the histograms for the two-strip center of
gravity (COG$_2$). This similarity suggested the
use of the amplitudes of the histograms directly
as weights of homogeneous  trackers (formed by identical
detectors). These apparently improbable attempts gave
results not too different from those of the more complex schematic
model. However, the absence of a global scaling factor, intrinsic to the schematic
model, imposes a limitation to homogeneous trackers
where the expressions of the parameter estimators are
invariant under a multiplicative constant. The
construction of super-lucky model eliminates
this limitation, maintaining the easiness of
the approach and a straightforward connection with the data.

\section{The COG$_2$ algorithm}

To obtain the PDF for the COG$_2$ algorithm, we have to define in
detail this algorithm. As previously recalled, the signals of
three strips must be simultaneously accounted: the strip with the
maximum signal (strip $\#\,2$ the seed strip) and the two lateral (strip $\# 1$ to
the right and strip $\# 3$ to the left). Any signal-value in these two strips
is accepted, also below the threshold for the insertion of
a strip-signal in a cluster by the cluster detection algorithm (as in~\cite{landi14}).
Around the strip $\#\,2$, the strip with the maximum
signal is selected between the two strips
$\#\,1$ and $\#\,3$. Due to the smallest number of strips,
this COG$_2$ has a very favorable signal-to-noise ratio.
It is the natural selection for orthogonal incidence on
strip detectors with strip widths near to
the average lateral drift of the primary charges~\cite{landi03}.
\subsection{The definition of the COG$_2$ algorithm}
The definition of COG$_2$ algorithm
is condensed in the following equation~\cite{landi06}:
\begin{equation}\label{eq:equation_7}
    x_{g2}=\frac{x_1}{x_1+x_2}\theta(x_1-x_3)-
    \frac{x_3}{x_3+x_2}\theta(x_3-x_1)\,.
\end{equation}
Where $x_1,\,x_2,\,x_3$ are the random signals in the three strips,
and $\theta(z)$ is the Heaviside $\theta$-function ($\theta(x)=0$
for $x\leq 0$ and $\theta(x)=1$ elsewhere). The two $\theta$-functions
select the lateral strip with the highest signal. No condition is
imposed on the strip $\#\,2$, although  it
has some constraints for its role of seed strip.
This choice eliminates inessential complications
and saves the normalization of the PDF.

Our aim is to reproduce
the gap for $x_{g_2}\approx\,0$, typical of the histograms of
COG$_2$ algorithm. This gap is given by the impossibility
(or lower probability)
to have $x_{g_2}\approx 0$ if the charge
drift populates one or both the two lateral strips.
The gap grows rapidly with an increase of these two charges.
The noise and our selection allow  forbidden values.

The $\eta$-algorithm of reference~\cite{belau} uses a slight different
definition. The term $-x_3/(x_2+x_3)$ of
equation~\ref{eq:equation_7} is modified
in $x_2/(x_2+x_3)$. In this way, the values of $\eta$
are contained in the interval
$\{0\vdash\dashv1\}$, and the gap for $x_{g2}\approx 0$ is spread
at the borders of this interval.

The constraints of equation~\ref{eq:equation_7},
on the three random signal $\{x_1,x_2,x_3\}$, are
inserted in the integral for the PDF of $  P_{x_{g2}}(x) $.
The integral expression is given by (with
the substitution of $x_{g_2}$ as $x$):
\begin{equation}\label{eq:equation_8}
\begin{aligned}
    P_{x_{g2}}(x)=&\int_{-\infty}^{+\infty}\mathrm{d}\,x_1\,
    \mathrm{d}\,x_2\, \mathrm{d}\,x_3  P_1(x_1)P_2(x_2)P_3(x_3)
    \Big[\delta\big(x-\frac{x_1}{x_1+x_2}\big)\theta(x_1-x_3)+\\
    &\delta\big(x+\frac{x_3}{x_3+x_2}\big)\theta(x_3-x_1)\Big] \,.
\end{aligned}
\end{equation}
The normalization of $P_{x_{g2}}(x)$ can be immediately proved with a direct
$x$-integration. The other integrals are executed with the
transformations: $x_1=\xi$, $x_1+x_2=z_1$,  $x_3=\beta$
and $x_3+x_2=z_2$. The jacobian of each couple of transformations
is one. The integrals on  $z_1$ and $z_2$ of the two $\delta$-functions
can be performed with the standard rules~\cite{landi06}.
The general form of $P_{x_{g2}}(x)$ for any type of
signal PDF $\{P_1,P_2,P_3\}$ becomes:
\begin{equation}\label{eq:equation_10}
\begin{aligned}
    P_{x_{g2}}(x)=\frac{1}{x^2}\Big[&\int_{-\infty}^{+\infty}\mathrm{d}\xi P_1(\xi)P_2(\xi\frac{1-x}{x})|\xi|\int_{-\infty}^{\xi}\mathrm{d}\beta P_3(\beta)+\\
    &\int_{-\infty}^{+\infty}\mathrm{d}\beta P_3(\beta)P_2(\beta\frac{-1-x}{x})|\beta|
    \int_{-\infty}^{\beta}\mathrm{d}\xi P_1(\xi)\Big] \,.
\end{aligned}
\end{equation}
Each strip has its own random additive noise uncorrelated
with that of any other strip. In strips without
MIP signals, the strip noise is well reproduced with
Gaussian PDFs. Thus, the PDFs for the signal plus
noise of the strip $i$ become:
\begin{equation}\label{eq:equation_2}
    P_i(z)=\frac{1}{\sqrt{2\pi}\,\sigma_i}
    \exp[-\frac{(z-a_i)^2}{2\sigma_i^2}]\ \ \ \ i=1,2,\cdots, n\,.
\end{equation}
The Gaussian mean values $\{a_i\}$ are the (noiseless)
charges collected by the strips and are positive numbers
(we assume to have signals from real particles). The
parameters $\{\sigma_i\}$ are the standard deviations of the
additive zero-average Gaussian noise.

The Gaussian PDFs of equation~\ref{eq:equation_2}, inserted in
equation~\ref{eq:equation_10}, allow the explicit expression of the two integrals
on $P_3(\beta)$ and $P_1(\xi)$ with the appropriate
erf-functions. Indicating the remaining integration variable  as $z$,
equation~\ref{eq:equation_10} becomes:
\begin{equation}\label{eq:equation_11}
\begin{aligned}
& P_{xg_2}(x)=\frac{1}{2\pi\sigma_1\sigma_2 x^2}\\
&\Big(\int_{-\infty}^{+\infty} \mathrm{d}
z \,|z|\Big\{\exp\big[-\frac{(z-a_1)^2}{2\sigma_1^2}-
\frac{(\frac{(1-x)z}{x}-a_2)^2}{2\sigma_2^2}\big]\,\frac{1}{2}
\big[1-\mathrm{erf}(\frac{a_3-z}{\sqrt{2}\sigma_3})\big]+\\
&\frac{\sigma_1}{\sigma_3}\exp\big[-\frac{(z-a_3)^2}{2\sigma_3^2}-
\frac{(\frac{(-1-x)z}{x}-a_2)^2}{2\sigma_2^2}\big]\,\frac{1}{2}
[1-\mathrm{erf}(\frac{a_1-z}{\sqrt{2}\sigma_1})]\Big\}\Big) \,.
\end{aligned}
\end{equation}
The combination of erf-functions and the $|z|$ render impossible
an analytical integration of equation~\ref{eq:equation_11}. The serial
development of the erf-function and its successive integration term by term is
too cumbersome to be of practical use. (A generalization
of the erf-function to the two dimensions is missing).  Thus, we
have to explore analytical approximations apt to be useful in maximum likelihood search.

\subsection{Small |x| approximation}

The small $|x|$ approximation is one of the easiest way
to handle equation~\ref{eq:equation_11}. The function
$P_2(z(1-x)/x)$ can be transformed to
approximate a Dirac $\delta$-function for small $|x|$:
\begin{equation}
\begin{aligned}
    &\frac{\exp\Big[-\big(\frac{1-x}{x}z-a_2\big)^2\frac{1}
    {2\sigma_2^2}\Big]}{\sqrt{2\pi} \sigma_2}=
    \frac{\exp\Big[-\big(\frac{z}{a_2}-\frac{x}{1-x}\big)^2
    \frac{(1-x)^2\,a_2^2}{2x^2\,\sigma_2^2} \Big]}
    {\Big(\frac{\,\sqrt{\,2\,\pi\,}\, \sigma_2\, |\,x\,|\,}{\,a_{\,2}\,|\,(1-x)\,|\,}\Big)}
    \frac{|x\,|}{a_2|(1-x)|}\\
    &\approx \frac{|x\,|}{a_2|(1-x)|}\delta(\zeta-\frac{x}{1-x})
    \ \ \ \ \ \ \ \zeta=\frac{z}{a_2}  \,.
\end{aligned}
\end{equation}
The effective standard deviation of the gaussian is $\sigma_2 |x|/(a_2 |1-x|)$,
this term, for $|x|\rightarrow 0$, allows to identify
the gaussian with a Dirac $\delta$-function. The term $|1-x|$ is useful to obtain
the combination $a_1/(a_1+a_2)$ in the exponent of the Gaussian-like function.
A similar transformation can be applied to $P_2(z(-1-x)/x)$, the integration on $\zeta$
is now immediate and the small $|x|$ probability $P_{xg_2}$ becomes~\cite{landi06}:
\begin{equation}\label{eq:equation_13}
\begin{aligned}
    P_{xg_2}(x)=&\frac{|a_2|}{2\sqrt{2\pi}}\Big\{\frac
    {\exp\big[-(x-\frac{a_1}{a_1+a_2})^2\frac{(a_1+a_2)^2}{2(\sigma_1^2(1-x)^2)}\big]
    \big(1-\mathrm{erf}\big[(\frac{a_3}{a_3+a_2}-x)\frac{a_2+a_3}
    {\sqrt{2}(1-x)\sigma_3}\big]\big)}{\sigma_1(1-x)^2}+\\
    &\frac{\exp\big[-(x+\frac{a_3}{a_3+a_2})^2\frac{(a_3+a_2)^2}{2(\sigma_3^2(1+x)^2)}\big]
    \big(1-\mathrm{erf}\big[(\frac{a_1}{a_1+a_2}+x)\frac{a_1+a_2}{\sqrt{2}(1+x)\sigma_1}\big]\big)}
    {\sigma_3\,(1+x)^2}\Big\} \,.
\end{aligned}
\end{equation}
The term $a_2$ is a positive constant
(the noiseless charge of the seed strip) and the absolute value
can be eliminated, but for future developments is
better to remember its presence.
Two different simulated distributions are reported
in~\cite{landi05,landi06} and compared with
equation~\ref{eq:equation_13}, the first one did not
contain the Landau fluctuations, the
second one contains approximate Landau fluctuations.
At orthogonal incidence, the Landau fluctuations are well
described by the fluctuations of the total collected charges.

The approximation of equation~\ref{eq:equation_13} reproduces, in a
reasonable way, the COG$_2$ PDF
also for non small $x$. In fact, the real useful range of $x$ is $|x|\leq 0.5$,
and the factor that is supposed small is $|x|\sigma_2/a_2$. But, the constant
$a_2$ is connected to seed
of the cluster and it has a high probability to be larger
than few times $\sigma_2$. The $a_2$ noisy detected part, $x_2$, must assure a
reasonable detection efficiency of the hit. Surely
equation~\ref{eq:equation_13} becomes useless around $x=\pm 1$.
In any case, better approximations are always useful, given that the probability
$P_{xg_2}(x)$ has to apply to a large set of experimental configurations.
A conceptual incompleteness of equation~\ref{eq:equation_13} is the lack of the
normalization. The normalization assures a constant probability of the impact
point, but its lack is not a real limitation
for the practical use of equation~\ref{eq:equation_13}.

\subsection{A better approximation for $P_{xg_2}(x)$}

A more accurate approximation for $P_{xg_2}(x)$ can be obtained
retaining the small $x$ approximation for the two
$\mathrm{erf}$-function of equation~\ref{eq:equation_11} and
integrating on $z$ the remaining parts~\cite{MATHEMATICA}. Now the two
integrals have analytical forms calculated in ref.~\cite{landi10}.
 This approximation saves also
the normalization.
\begin{equation}\label{eq:equation_17}
    \begin{aligned}
    P_{xg_2}(x)=&\Big\{\frac{\Big|a_2(1-x)\sigma_1^2+a_1x\sigma_2^2\Big|}
    {2\sqrt{2\pi}[(1-x)^2\sigma_1^2+x^2\sigma_2^2]^{3/2}}\exp\big[-
    (\frac{a_1}{a_1+a_2}-x)^2\frac{(a_1+a_2)^2}{2(\sigma_1^2(1-x)^2+x^2\sigma_2^2)}\big]\Big\}\\
    &\Big\{1-\mathrm{erf}\big[(\frac{a_3}{a_3+a_2}-x)\frac{a_2+a_3}
    {\sqrt{2}(1-x)\sigma_3}\big]\Big\}+ \  \ a_1\leftrightarrow a_3,
    \ \sigma_1\leftrightarrow\sigma_3, \ \ x\rightarrow -x\,.
    \end{aligned}
\end{equation}

\noindent
Although equation~\ref{eq:equation_17} represents a better approximation
compared to equation~\ref{eq:equation_13}, some discrepancies remain in very
special configurations. These  discrepancies are absent, or heavily reduced,
in the longer approximation reported at the end of ref.~\cite{landi10}.
However, for our necessity this approximation suffices.

\section{The lucky model}

A set of simulations highlight some peculiar properties of the
PDF of equation~\ref{eq:equation_17}. The effective use of
equation~\ref{eq:equation_17} in track reconstruction,
and similarly for all the other PDFs, requires the completion
with the functional dependence from the hit impact point.
The method for this completion is discussed in~\cite{landi05} with
the construction of the functions $a_j(\varepsilon)$. Differently
from their definition in the previous equations,
the $a_j(\varepsilon)$ of~\cite{landi05} are constructed normalized to one.
For tuning them to each hit, the parameter $E_0$ scales the $a_j$
to the appropriate values; $E_0$ is
the noiseless total charge collected
by the three strips of each hit.
The functions $a_j(\varepsilon)$ are obtained by large
averages of hit positioning given by the
$\eta$-algorithm~\cite{landi03}. The data of
the test-beam~\cite{pamela} were used for those
constructions. The unbiasedness of the
$\eta$-algorithm and the special averages~\cite{landi05} give to
the $a_j$  realistic functional dependencies from the
hit impact point $\varepsilon$.
The finiteness of the averages leave small artifacts,
but they have no detectable effects on the distributions of
the fitted parameters.
\begin{figure} [h!]
\begin{center}
\includegraphics[scale=0.45]{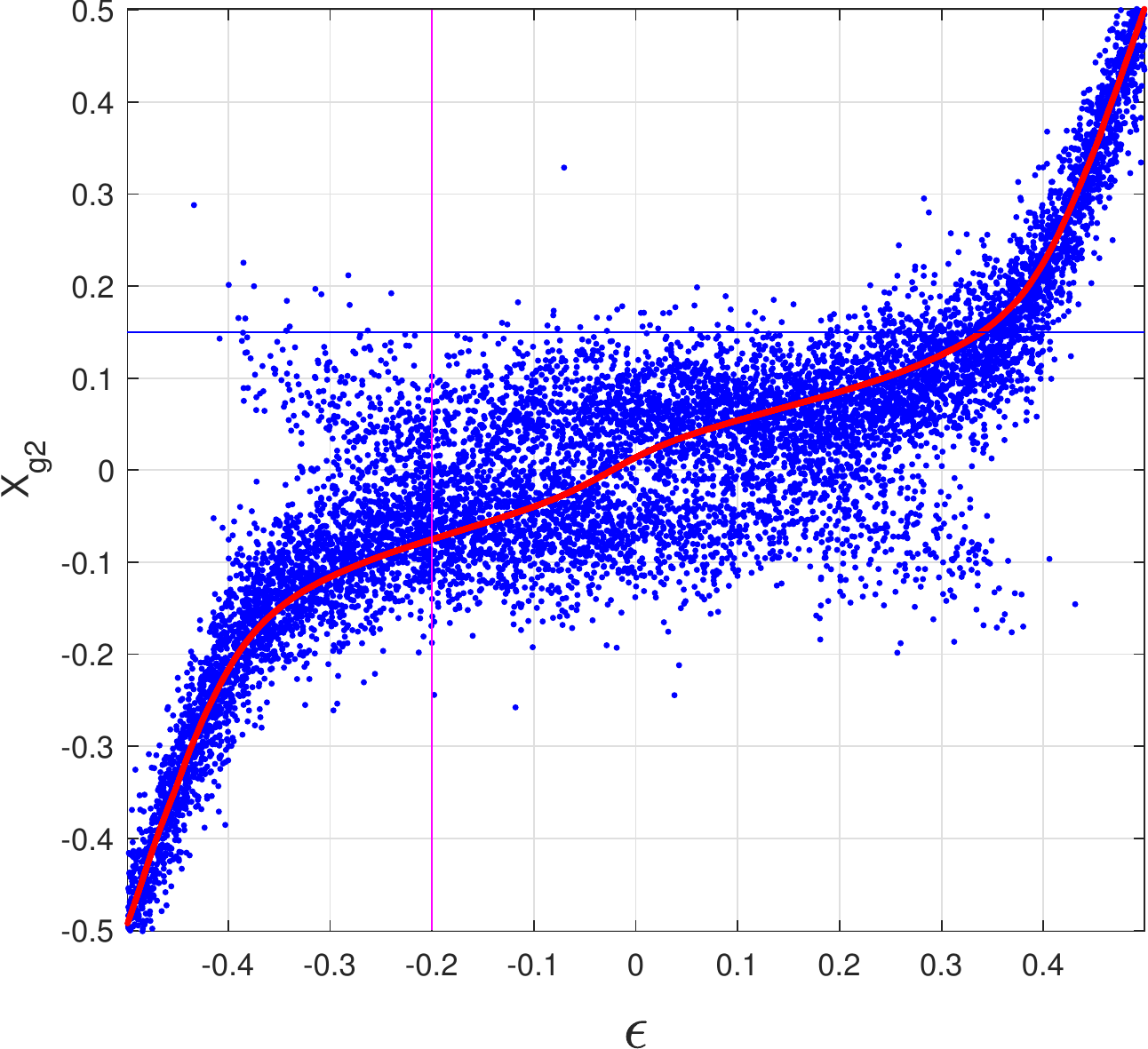}
\includegraphics[scale=0.5]{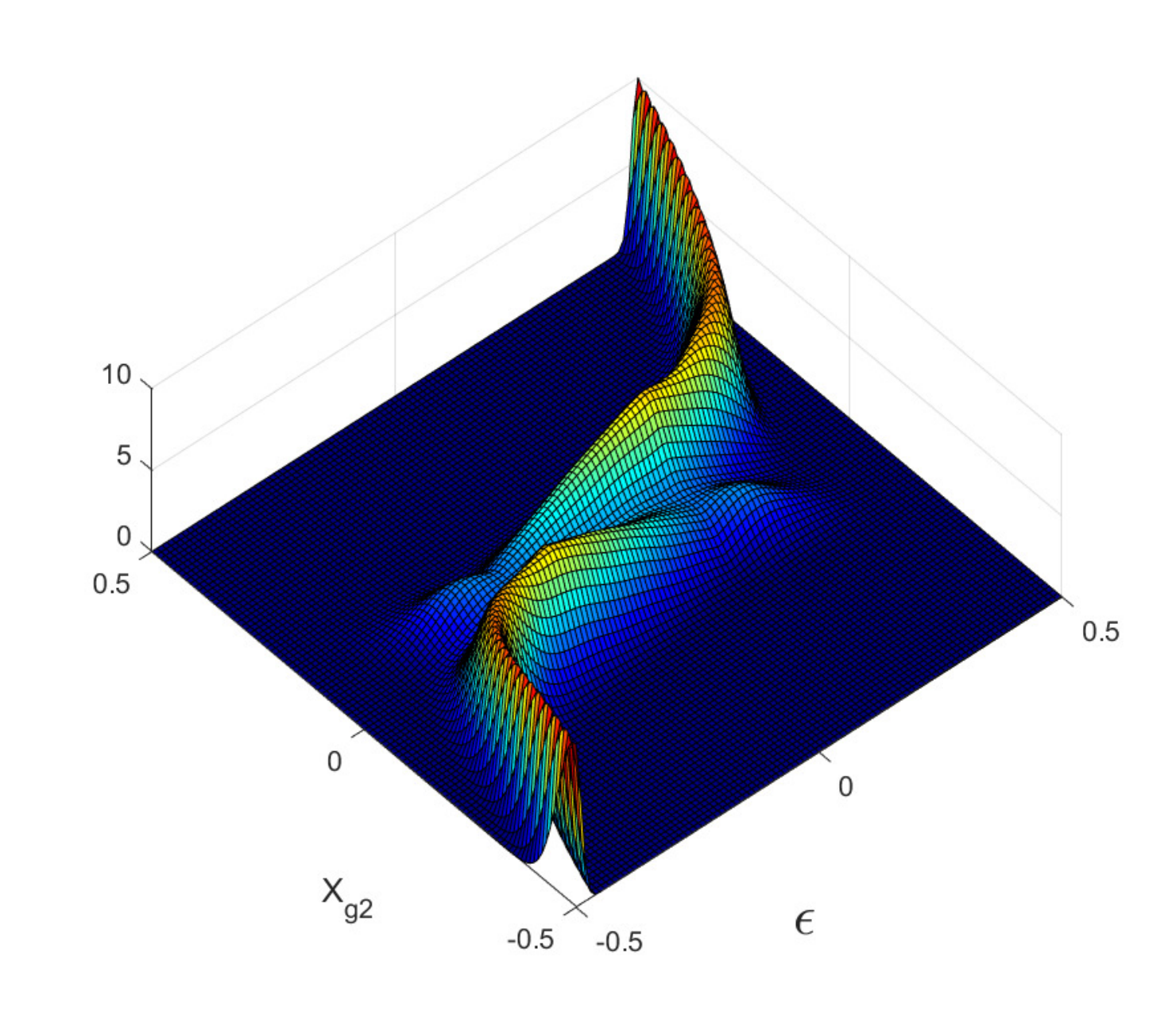}
\caption{Left side. Scatter-plot of a sample of simulated
$x_{g2}$ data in function of the impact point $\varepsilon$,
the red line is the $\eta$-algorithm.
The magenta line and the blue line are for the
following figures. The right side is a 2D illustration of the
PDF for $x_{g2}$ in function of the impact point $\varepsilon$
for $E_0=150$ ADC counts.}
\label{fig:figure_1}
\end{center}
\end{figure}

\begin{figure} [h!]
\begin{center}
\includegraphics[scale=0.41]{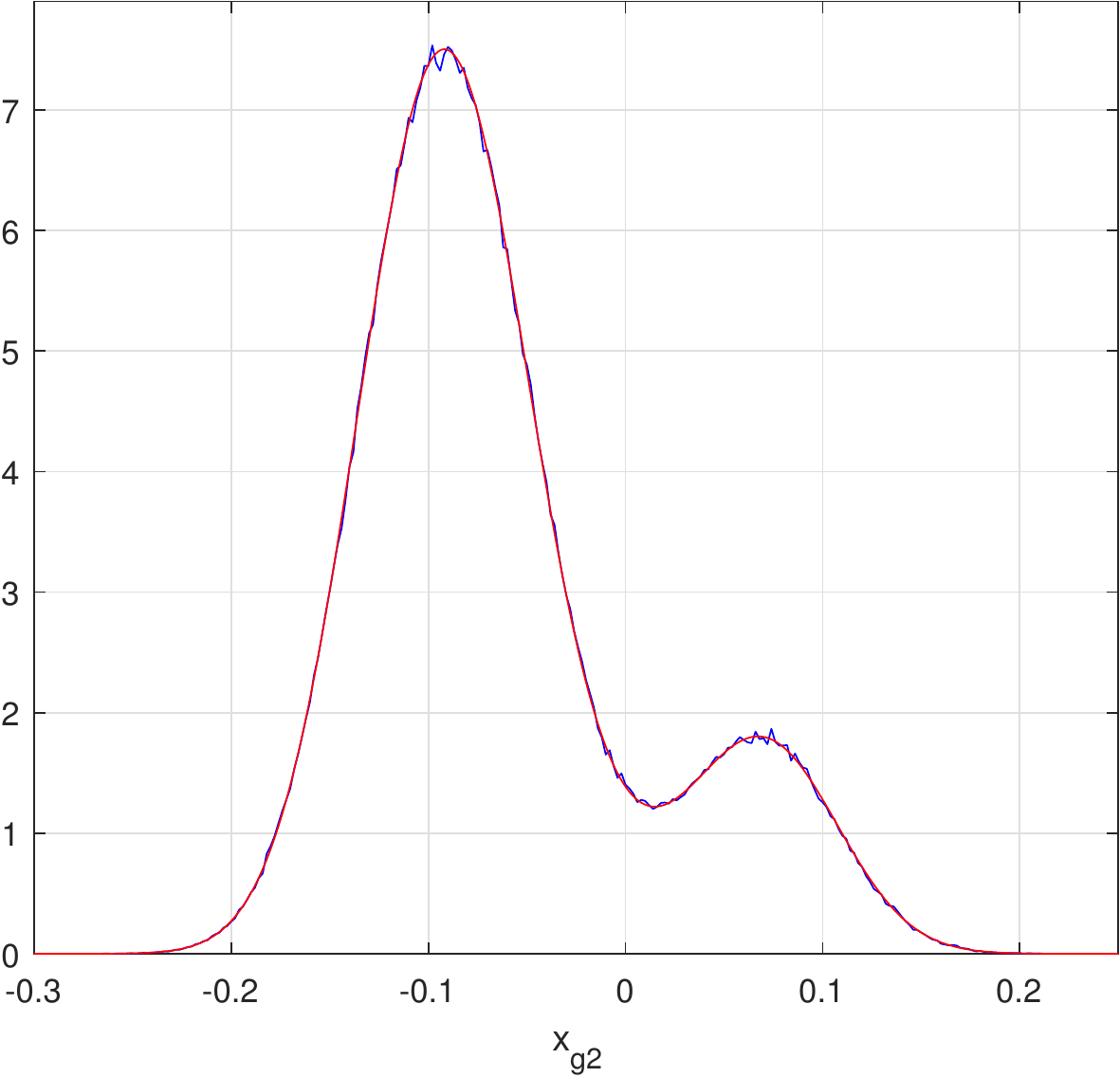}
\includegraphics[scale=0.40]{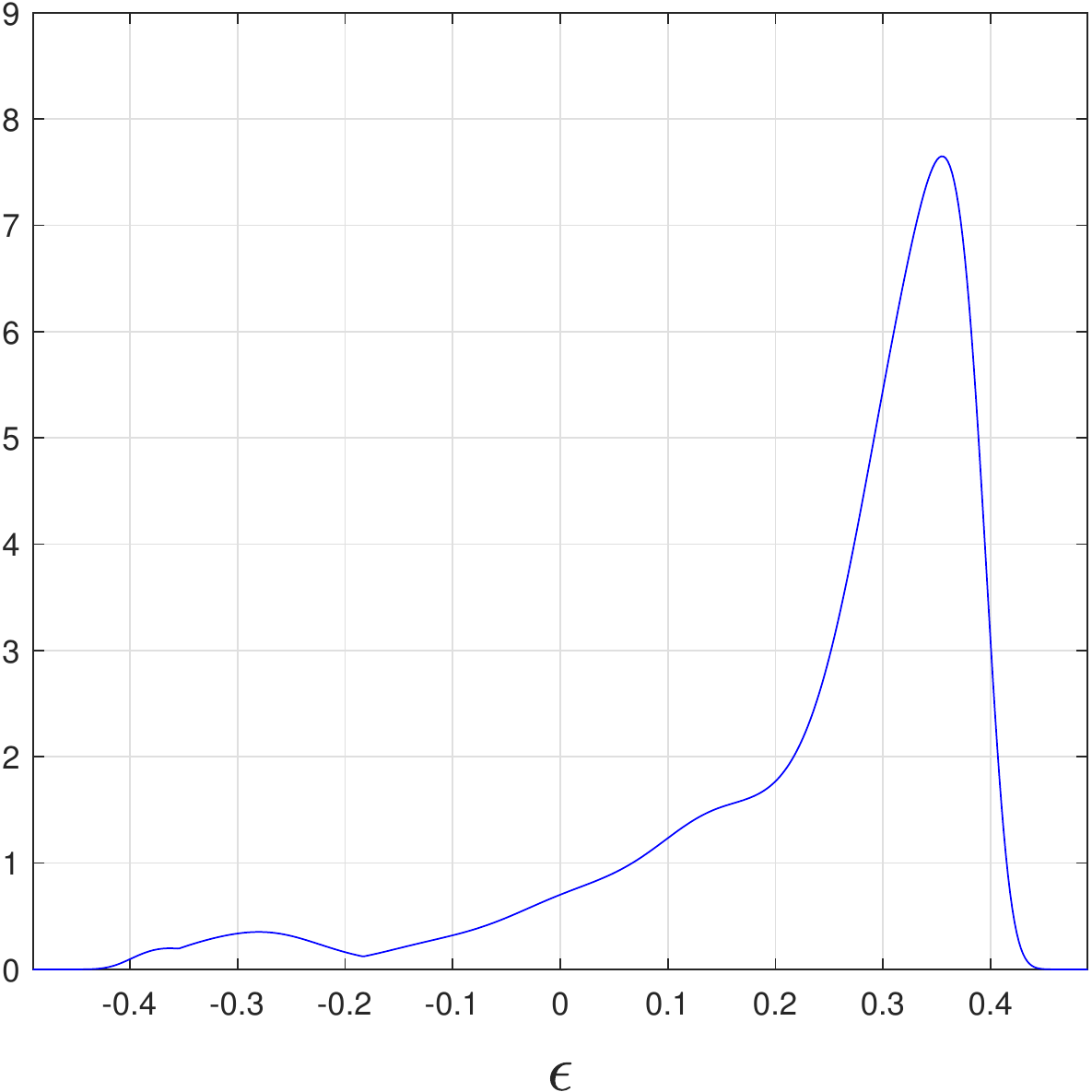}
\caption{ Left side: Empirical PDFs of $x_{g_2}$ (blue line)
compared with equation~\ref{eq:equation_17} (red line) for a
model of silicon detector~\cite{landi05} for an impact
point $\varepsilon=-0.2$, all the $\sigma=8$ ADC counts, and $a_1=0.01\,E_0$,
$a_2=0.91\,E_0$ and $a_3=0.08\,E_0$ are the charges collected by the three
strips, here $E_0$, the total charge of the three strips, is 150 ADC counts
(the magenta line of figure~\ref{fig:figure_1}). The right side is the
probability distribution for $x_{g2}=0.15$, the blue line in the
scatter-plot of figure~\ref{fig:figure_1}}
\label{fig:figure_2}
\end{center}
\end{figure}

A simple simulation can be used to verify the
consistency of equation~\ref{eq:equation_17}
and to illustrate the weak gap present  in a
distribution of simulated $x_{g_2}$ in the
scatter-plot of figure~\ref{fig:figure_1}.
The data of the simulation are generated
with the function {\tt randn} of MATLAB~\cite{matlab} and with
the equations $x_i=\sigma_1$\,{\tt randn(1,N)}$+a_i*E_0$,
and inserted in equation~\ref{eq:equation_7} to
calculate a large sample of $x_{g2}$.
Values for $a_i$ ($a_1=0.01$, $a_2=0.91$, $a_3=0.08$,
$E_0=150$ ADC counts and $\sigma_1=8$ ADC counts),
$\sigma_i$ are extracted from~\cite{landi05} for
orthogonal incidence on the two types of silicon
detectors studied there. The type of detector
considered in figures~\ref{fig:figure_1}
and~\ref{fig:figure_2}
is of normal type with a high noise of 8 ADC
counts, one of the two types studied in~\cite{landi05,landi06}. To simplify
the simulations, all the $\sigma_i$ are identical to
the most probable noise of the strips.
The noisy strips are very few, and it is
an useless complication to explore different
$\sigma_i$. The left side of figure~\ref{fig:figure_2}
reports the overlaps of the analytical PDF
and the PDF obtained from the simulated data.
The probability decrease  between the principal and
secondary maximum of figure~\ref{fig:figure_2},
originating a similar reduction along the magenta
line in the scatter-plot of figure~\ref{fig:figure_1}.
The secondary maximum is produced by the noise that
promotes the minority noiseless signal to becomes
the greater one. Signal clusters with lower total
charge show larger gaps.

The right side of figure~\ref{fig:figure_2} shows
the probability distribution of the hit impact
points for a given value of the two strip COG
$x_{g2}$. The $x_{g2}=0.15$ is the constant value of $x_{g2}$,
the blue line in the scatter-plot of figure~\ref{fig:figure_1}.
The integral of this distribution gives the probability
for this value of $x_{g2}$ for 150 ADC counts. We limit the integration
to a two strip length centered to the maximum of the
distribution. In this way, the histogram of $x_{g2}$
is analytically reconstructed and reproduces well the
histogram of the simulated data~\cite{landi05}. Products of probability
distributions, similar to that of figure~\ref{fig:figure_2},
are used to find the maximum likelihood of a set of hits
for a track. This figure shows also the presence of
outliers in the tail of the distribution. These outliers
are difficult to handle because they are masked as good
hits in the schematic model. Instead the maximum
likelihood is able to avoid their disturbances in
the fits. In~\cite{landi05}, we illustrate one of
the worst outlier and how the maximum likelihood find
almost exact parameters of the fitted track. We
accumulated many similar simulated events with
excellent maximum likelihood reconstructions and worst
schematic model and standard fit reconstructions.
\begin{figure} [h!]
\begin{center}
\includegraphics[scale=0.47]{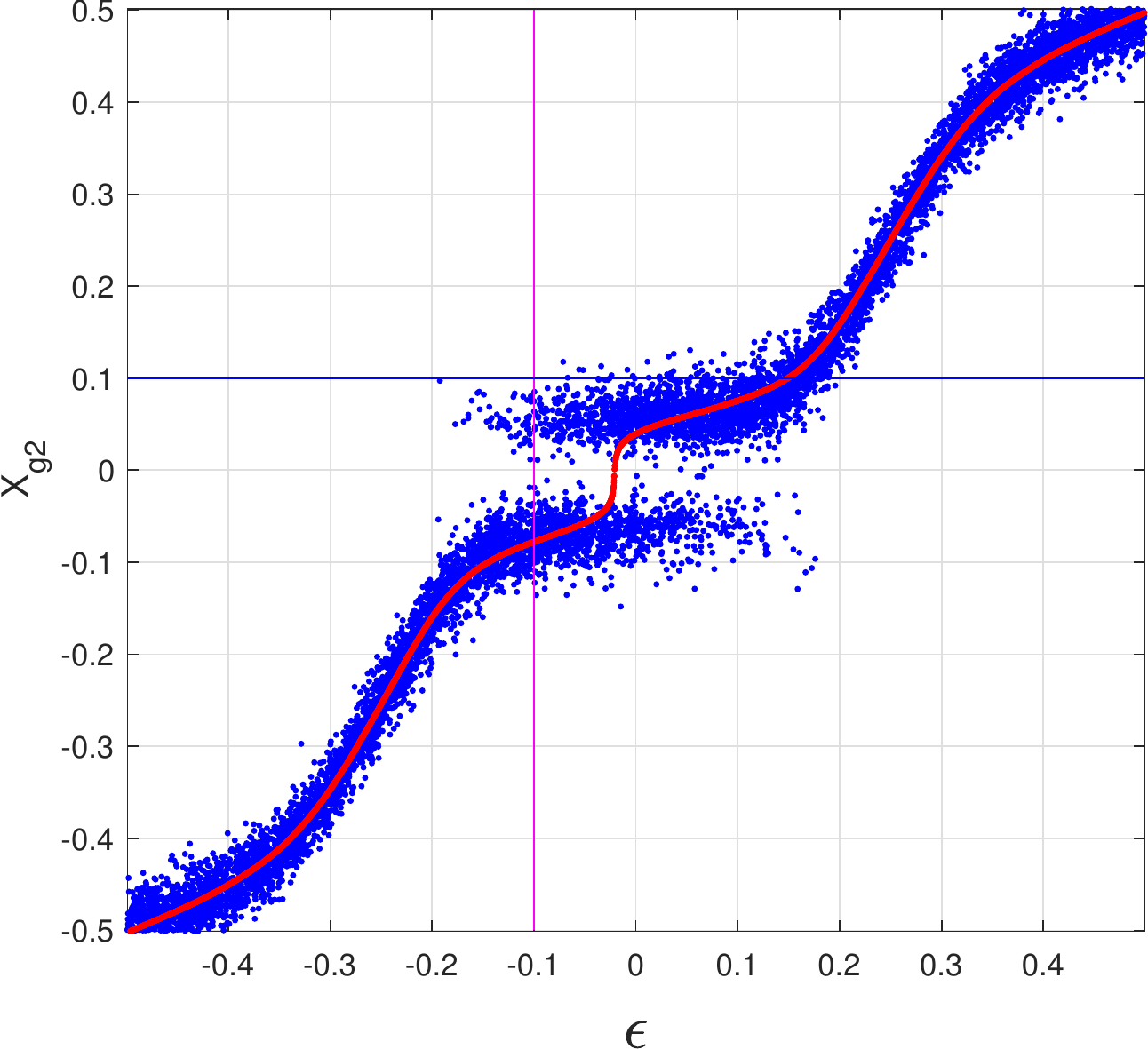}
\includegraphics[scale=0.50]{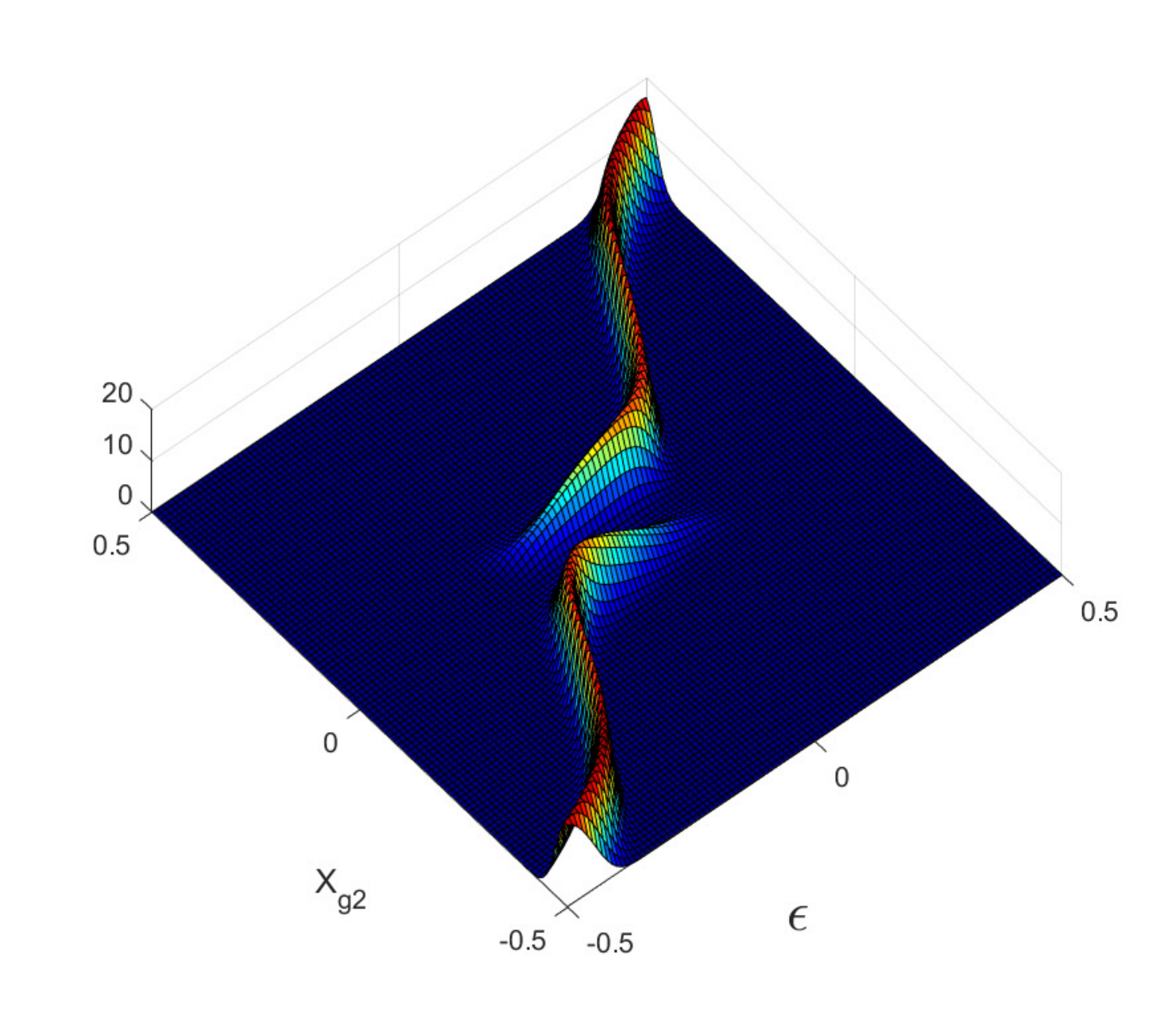}
\caption{ Similar to figure~\ref{fig:figure_1} for floating strip detectors.
Left side. Scatter-plot of $x_{g2}$ data in function
of $\varepsilon$. The right side.  2D  PDF
for $x_{g2}$ in function $\varepsilon$.   }
\label{fig:figure_1a}
\end{center}
\end{figure}
\begin{figure} [h!]
\begin{center}
\includegraphics[scale=0.47]{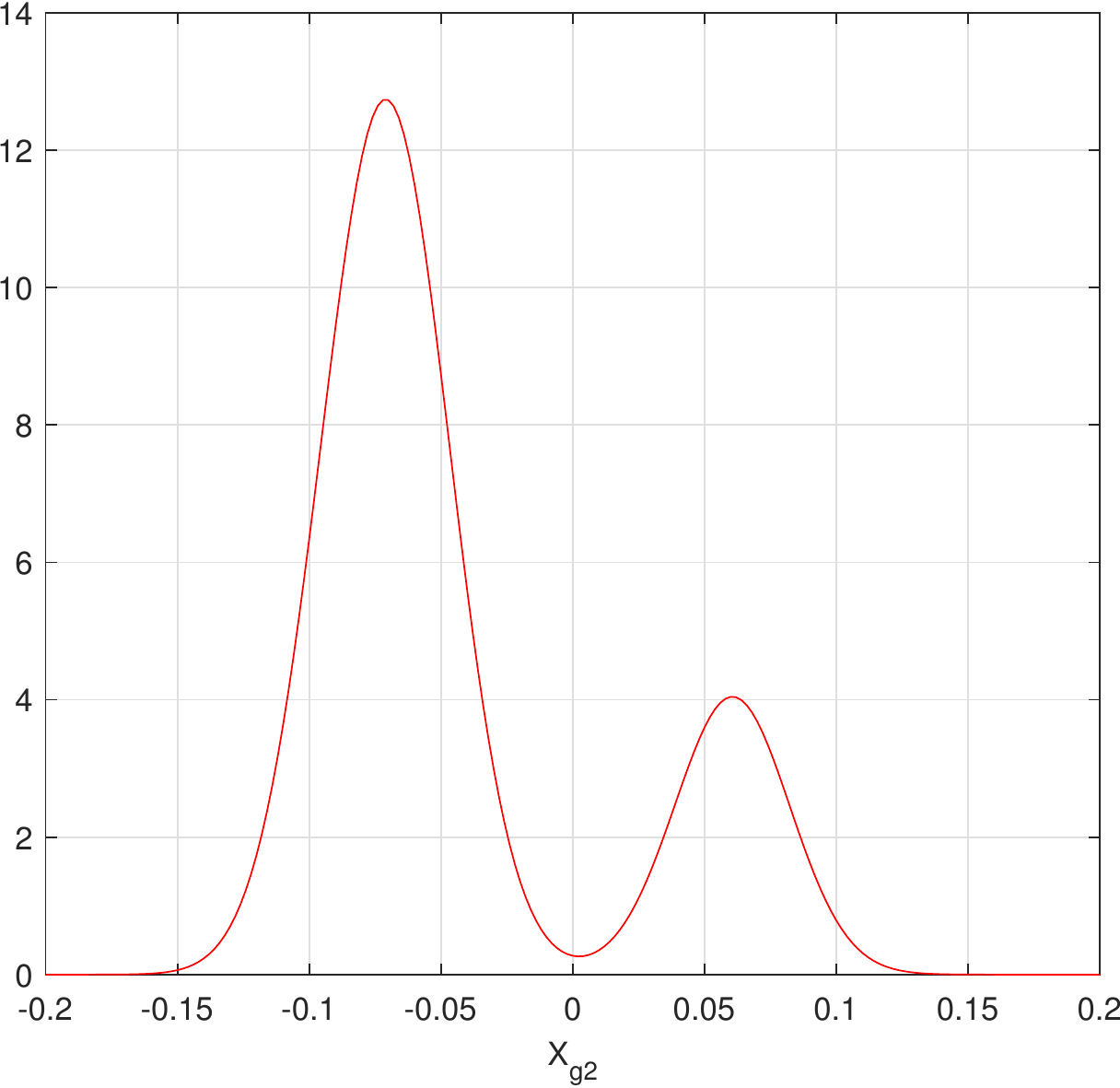}
\includegraphics[scale=0.47]{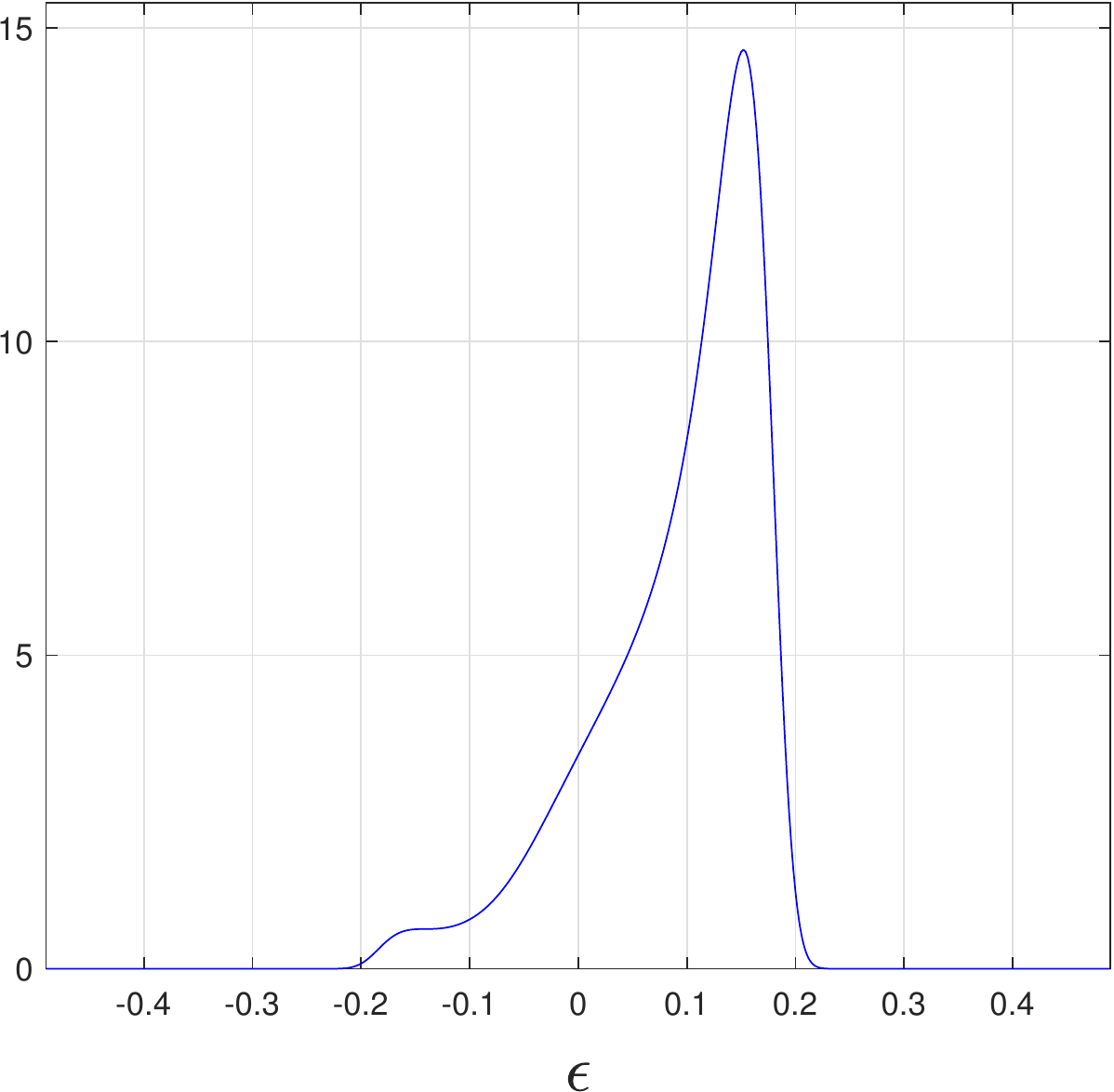}
\caption{ Floating strip detector. Left side: analytical  PDFs of $x_{g_2}$
for $\varepsilon=-0.1$ and E$_0=150$ ADC counts. Right side: the
probability distribution for $x_{g2}=0.1$, the blue line in the
scatter-plot of figure~\ref{fig:figure_1a}}
\label{fig:figure_2a}
\end{center}
\end{figure}
Plots similar to figure~\ref{fig:figure_1} can be produced
for the floating strip detector~\cite{landi05,pamela}. The plots
illustrate the large differences of this excellent detector type
with very low noise and the floating
strips, able to distribute the incoming signal to the nearby strips.
The gap for COG$_2\approx\, 0$ is a real gap without data in a scatter-plot
with a moderate number of events. The effects of the low noise (4 ADC)
are quite evident in all the distributions.

\subsection{A demonstration of the lucky model}

The availability of the analytical equations for
the COG PDFs allows a direct discussion of the
lucky model~\cite{landi15}.
The structure of~\cite{landi15} had to be heavily
deviated from its aims to complete this discussion.
As sketchily illustrated in~\cite{landi15}, this sub-optimal
model is suggested by the similarity of the trends
of the effective standard deviations for the schematic model
and the trends of the $x_{g2}$ histograms (similarly for
the COG$_3$ histograms). Those effective standard deviations
were obtained from the variances of
functions of $\varepsilon$, two of them illustrated
in the right side of figure~\ref{fig:figure_2} and
figure~\ref{fig:figure_2a}, as explained
in~\cite{landi05,landi06}. The scatter-plots
of these effective standard deviations clearly show
the trends of the $x_{g2}$ histograms.
In~\cite{landi05}, we gave an approximate motivation of
this non obvious correlation. We can complete those motivations
with more details. The reasoning of~\cite{landi05} suppose
(as always) an uniform population of events on a strip; thus,
for regions with large effective variances, a relative larger
fraction of events ends up to the corresponding $x_{g2}$-values.
Instead, for small effective variances, a relative lower fraction
of events ends up to these $x_{g2}$-values. Assuming the
correctness of this explanation, a test of its effectiveness
in improving a fit is a natural output. Although the
result of this "lucky" test of~\cite{landi15} was
successful, a detailed supporting proof is essential
for a confident use. This proof was
a pledge of reference~\cite{landi15}.
However, the construction of this proof requires an
analysis  of equation~\ref{eq:equation_17} and
the left sides of figures~\ref{fig:figure_2}
and~\ref{fig:figure_2a}, tools not available
in~\cite{landi15}.

\begin{figure} [h!]
\begin{center}
\includegraphics[scale=0.41]{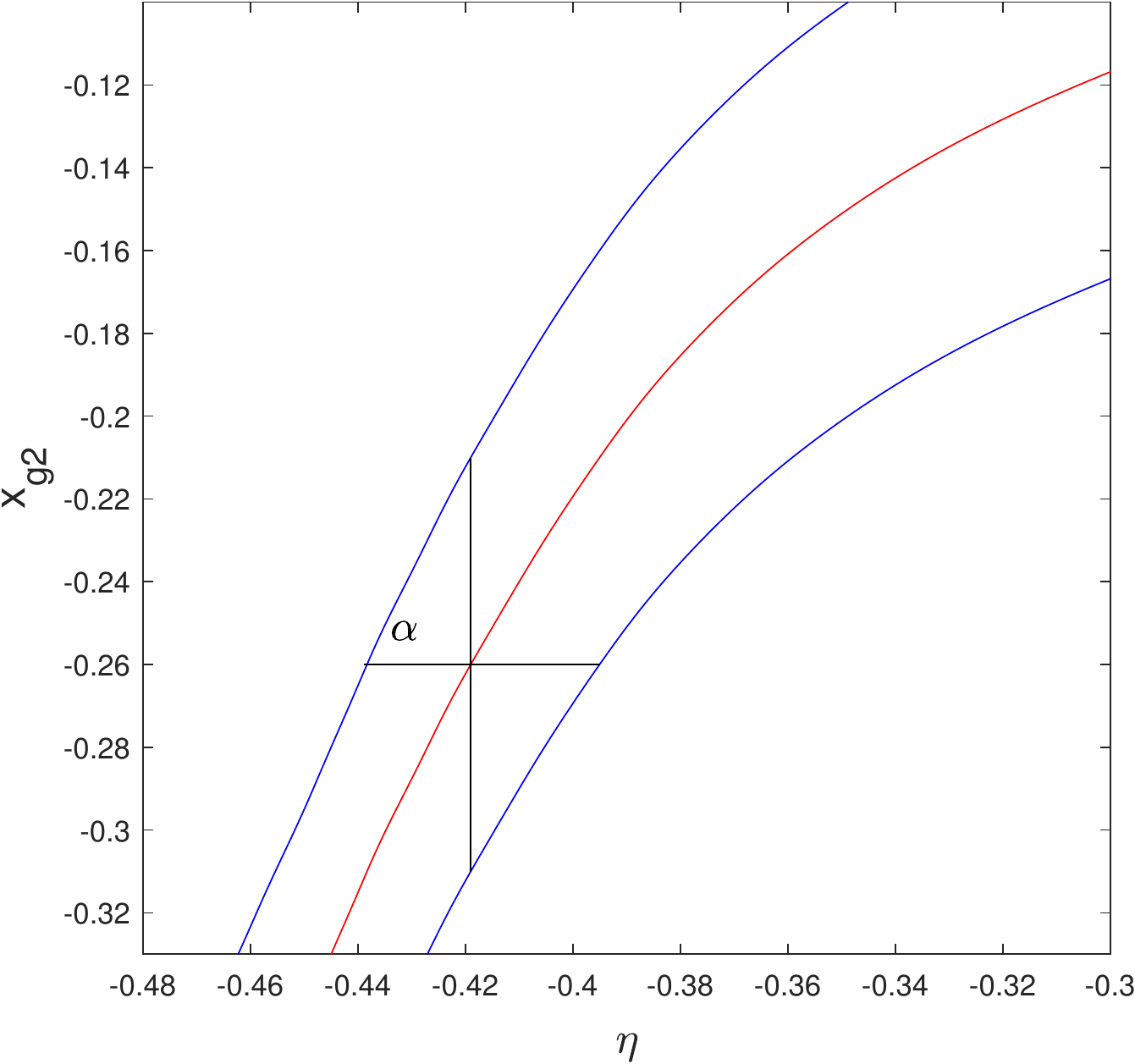}
\includegraphics[scale=0.41]{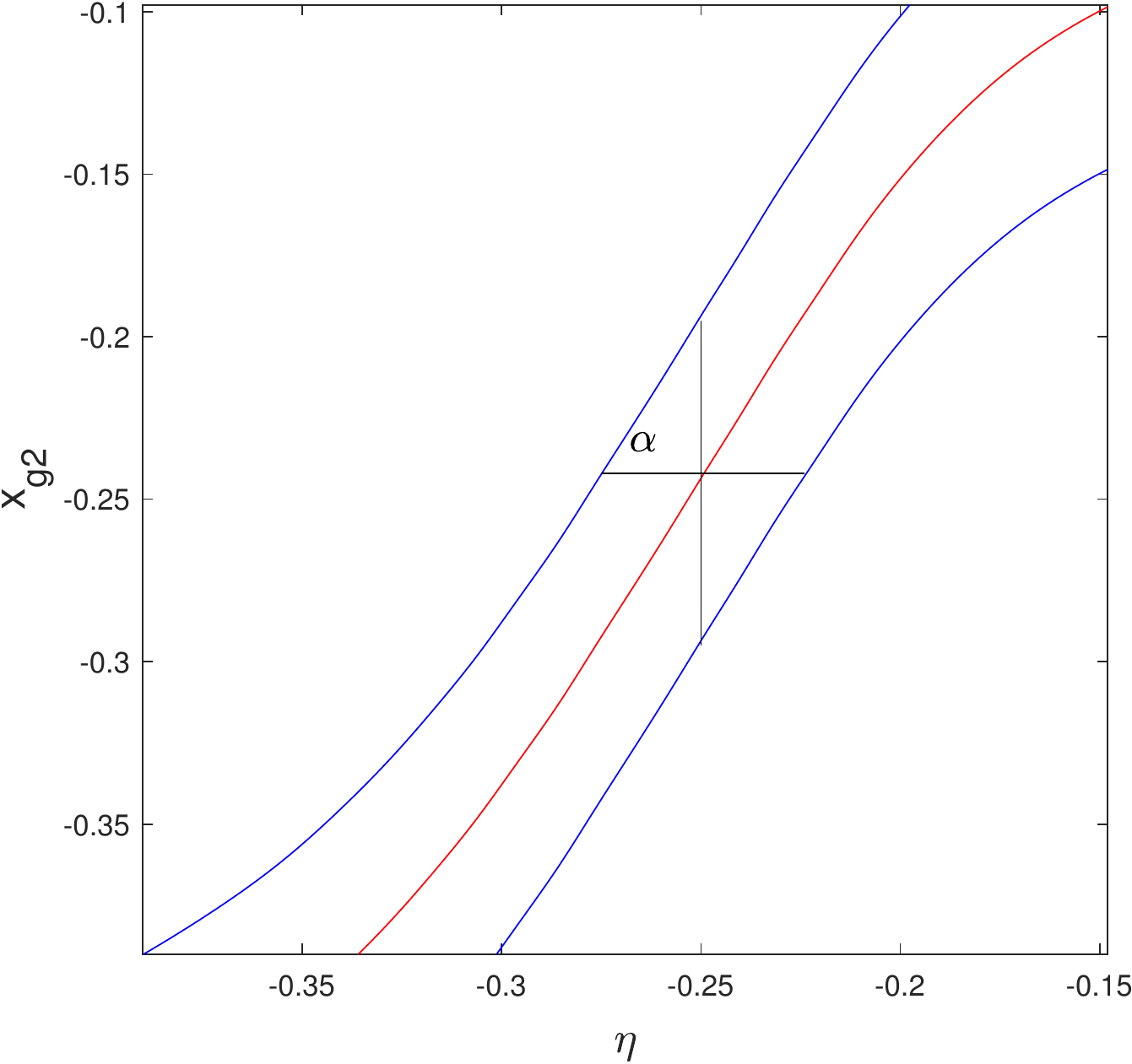}
\caption{ Relation of the amplitude of the $x_{g2}$
error of equation~\ref{eq:equation_17} with the positioning error
of a given value of COG$_2$ due to an observation.
Left side: normal strip detector. Right side: floating strip detector}
\label{fig:figure_3}
\end{center}
\end{figure}

Equation~\ref{eq:equation_17} shows the $P_{xg_2}(x)$
as formed by two bumps (approximatively Gaussian) whose
maxima
follows the noiseless function $x_{g2}(\varepsilon)$. Beyond the center
of the strip, the last part of each
bump decreases rapidly as illustrated in the 2D
PDF of figure~\ref{fig:figure_1}
and figure~\ref{fig:figure_1a}.
As discussed in~\cite{landi03}, the $\eta$-algorithm produces
functions that show  strong similarity with
the noiseless $x_{g2}(\varepsilon)$,
in the relevant parts of the two bumps.
Therefore, we can use the functions $x_{g2}(\eta)$, of the
$\eta$-algorithm, in place of the exact noiseless
$x_{g2}(\varepsilon)$, and to follow the path
of the bump maxima and two corresponding positions at the
half-maximum. The distance between these two positions is the
full-width-at-half-maximum of the $P_{xg_2}(x)$ in the direction
of the hit impact point.  Each one of these two points
follow paths parallel to the noiseless $x_{g2}(\varepsilon)$
one above and the other below the path followed by the bump maximum.
Again, we approximate these paths with the $x_{g2}(\eta)$-function.
The two sides of figure~\ref{fig:figure_3}(out of scale
for a better illustration) report these paths
for the two types of detectors of~\cite{landi05,landi06}.
The vertical segments are the
full-width-at-half-maximum of $P_{xg_2}(x)$
and will be defined as $\sigma_{x_{g2}}$.
The positioning errors $\sigma_\eta$ of a generic $x_{g2}^0$,
possible realizations of their COG$_2$,
are the horizontal black lines.
The amplitudes of the horizontal segments are
the positioning errors relevant for the weighted least
squares.
The blue lines
are curved lines and only approximatively describe a
triangle, the ratio of $\sigma_{x_{g2}}$ and $\sigma_\eta$
can be estimated as:
\begin{equation}\label{eq:lucky_model}
    \tan(\alpha)\approx\frac{\sigma_{x_{g2}}}{\sigma_\eta} \ \ \ \
    \frac{1}{\tan(\alpha)}=\frac{\mathrm{d}\,\eta}{\mathrm{d}
    \,x_{g2}}\big|_{x_{g2}^0}=\Gamma(x_{g2}^0)\ \ \ \ \
    \sigma_\eta\approx\frac{\sigma_{x_{g2}}}{\tan(\alpha)}
    =\sigma_{x_{g2}}\,\Gamma(x_{g2}^0)
\end{equation}
where $\Gamma(x_{g2}^0)$ is the amplitude of the normalized
histogram of COG$_2$ for the value $x_{g2}^0$. In fact,
the starting point of the $\eta$-algorithm of~\cite{landi03} is the
differential equation ${\mathrm{d}\,\eta}/{\mathrm{d}\,x_{g2}}=\Gamma(x_{g2})$.
Neglecting the differences of $\sigma_{x_{g2}}$, the $\eta$-algorithm gives all
the elements for a good fit; the corrections of the COG$_2$ systematic errors and
the weight for track fitting in an array of identical detector layers.
In this case, the expressions of the parameter estimators of
the fit are independent from the (assumed) constant
$\sigma_{x_{g2}}$. Therefore, the values of $\Gamma(x_{g2})$
can be used directly as weights of the observations
in a weighted least squares of a track.

In presence of large data gaps (i.e. absence of data)
in the histogram of COG$_2$, the function $x_{g2}(\eta)$
acquires discontinuities that complicate the plots
of figure~\ref{fig:figure_3}, as discussed in~\cite{landi01}.
In any case, the presence of large gaps in the
COG histograms should be avoided for the excessive
loss of information. It is better the use of
COG-algorithms with more strips.

\subsection{Advanced form for the lucky model: The super-lucky model }

The precedent discussion is a justification of the
simple lucky model of~\cite{landi15}. We
utilized this model with a set of identical
detectors, inserting directly the amplitudes
of the COG$_2$ histograms in place of effective
standard deviations of the observations.
The scaling factor ($\sigma(x_{g2})=$constant)
of the this approximate guess is simultaneously
present in the numerator and in the denominator
of the expressions of the fitted parameters and
it simplifies. This simplification is impossible
for arrays with detectors of different type.
Hence, the (assumed) constant vales  of the
scaling factors ($\sigma_{x_{g2}}$) become relevant
to tuning the amplitudes of different
histograms with the properties of the different
detectors. The full calculation of some effective
standard deviations of the schematic model looks
unavoidable.
However, the demonstration of equation~\ref{eq:lucky_model}
recalls the attention to the possible variations of
$\sigma_{x_{g2}}$. We supposed that the principal
bumps of equation~\ref{eq:equation_17} are very near
to Gaussian PDFs. We can push  this assumption
forward and extract from equation~\ref{eq:equation_17}
approximate forms for $\sigma_{x_{g2}}$ (abandoning the
full-width-at-half-maximum in favor of the usual
standard deviation):
\begin{equation}
\begin{aligned}
    &\sigma_{x_{g2}}^R=\frac{\sqrt{\sigma_1^2(1-|X^R|)^2+
    \sigma_2^2\,(X^R)^2}}{a_1+a_2} \ \ \ \ \ X^R=\frac{a_1}{a_1+a_2}\\
    &\sigma_{x_{g2}}^L=\frac{\sqrt{\sigma_3^2(1-|X^L|)^2+\sigma_2^2
    \,(X^L)^2}}{a_3+a_2} \ \ \ \ \
    X^L=-\frac{a_3}{a_3+a_2}
\end{aligned}
\end{equation}
The substitutions of $X^R$ and $X^L$, systematically in
equation~\ref{eq:equation_17}, transform the two bumps in two
Gaussian PDFs. In any case, also these approximate forms are
outside our reach, $a_1$, $a_2$ and $a_3$ are the noiseless
signals released by the MIP. Instead, we have their noisy
version. Similarly for $X^R$ and $X^L$ that are the noiseless
COG$_2$, ratio of noiseless signals. The only well defined
parameters are $\sigma_1$, $\sigma_2$ and $\sigma_3$, the
strip noises, calculated at the initialization stage of the
strip detectors.
However, we can try to combine the noisy data to see what
happen (we were lucky with the COG$_2$ histogram in~\cite{landi15}).
The COG$_2$ algorithm is described in equation~\ref{eq:equation_7}
with the definition of $x_{g2}$, therefore, the approximate
(super-lucky) $\Sigma_{sup}$ could be:
\begin{equation}\label{eq:sigma_lucky}
\begin{aligned}
    &\Sigma_{sup}=\frac{\sqrt{\sigma_1^2(1-|x^R|)^2+
    \sigma_2^2\,(x^R)^2}}{x_1+x_2}\,\theta(x_1-x_3)+
    \frac{\sqrt{\sigma_3^2(1-|x^L|)^2+\sigma_2^2\,(x^L)^2}}{x_3+x_2}\,\theta(x_3-x_1)\\
    &x^R=\frac{x_1}{x_1+x_2} \ \ \ \ \ \ \ \ \ x^L=-\frac{x_3}{x_3+x_2} \ \ \ \ \ \ \ \ \ \ \ \ \ \ \ \ \sigma_\eta=\Sigma_{sup}(x_{g2})\,\Gamma(x_{g2})
\end{aligned}
\end{equation}
The insertion of equation~\ref{eq:sigma_lucky} in the calculation of the effective
standard deviations of the observation can be compared with those of the
schematic model for the two widely different detector types of our simulations.
We tested also the use of the soft cut-offs with the erf-functions of
equation~\ref{eq:equation_17} in place of the sharp cut-offs of
equation~\ref{eq:sigma_lucky}, but this insertion adds complications
without any visible effect on the simulations.
\begin{figure} [h!]
\begin{center}
\includegraphics[scale=0.4]{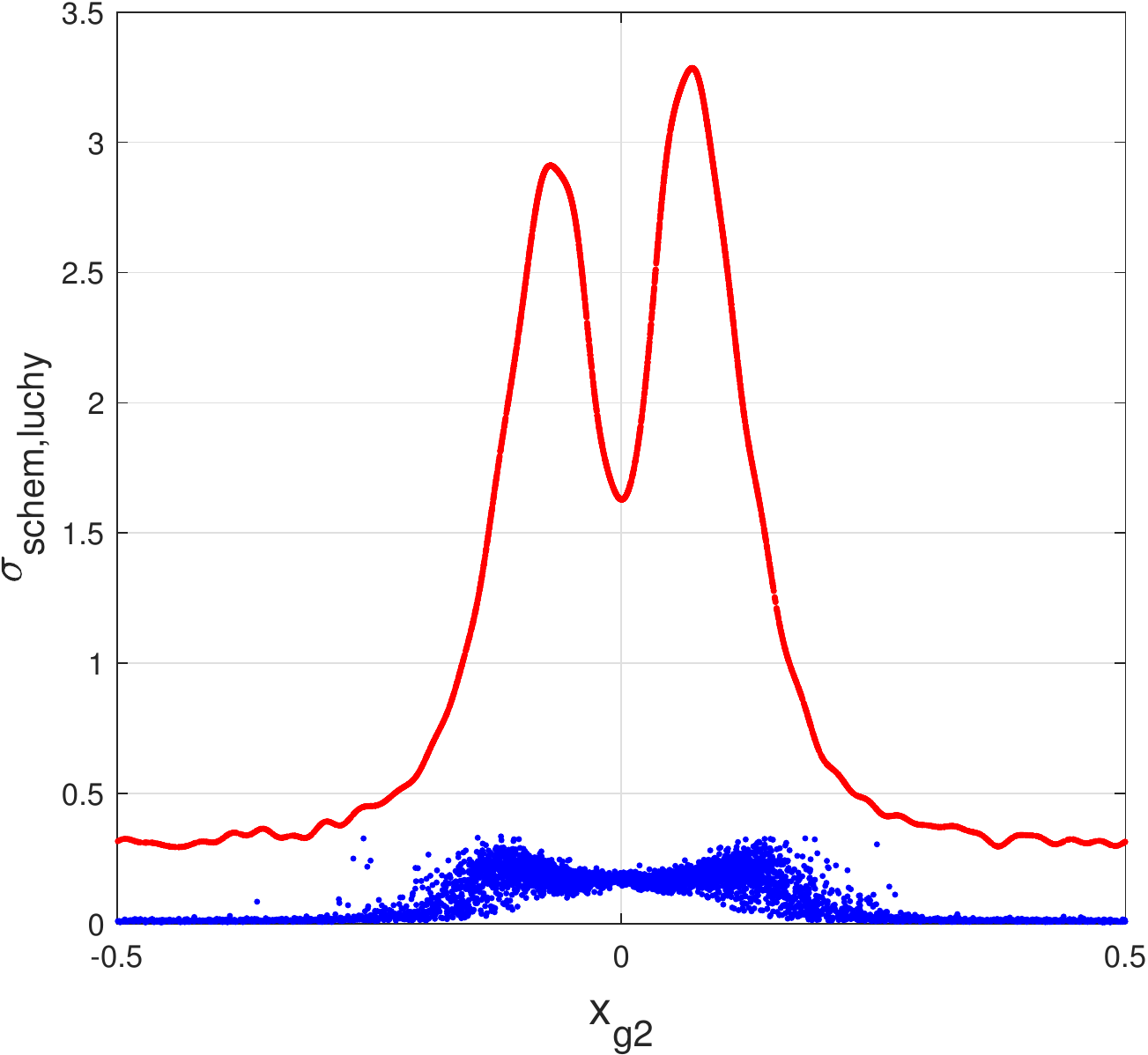}
\includegraphics[scale=0.4]{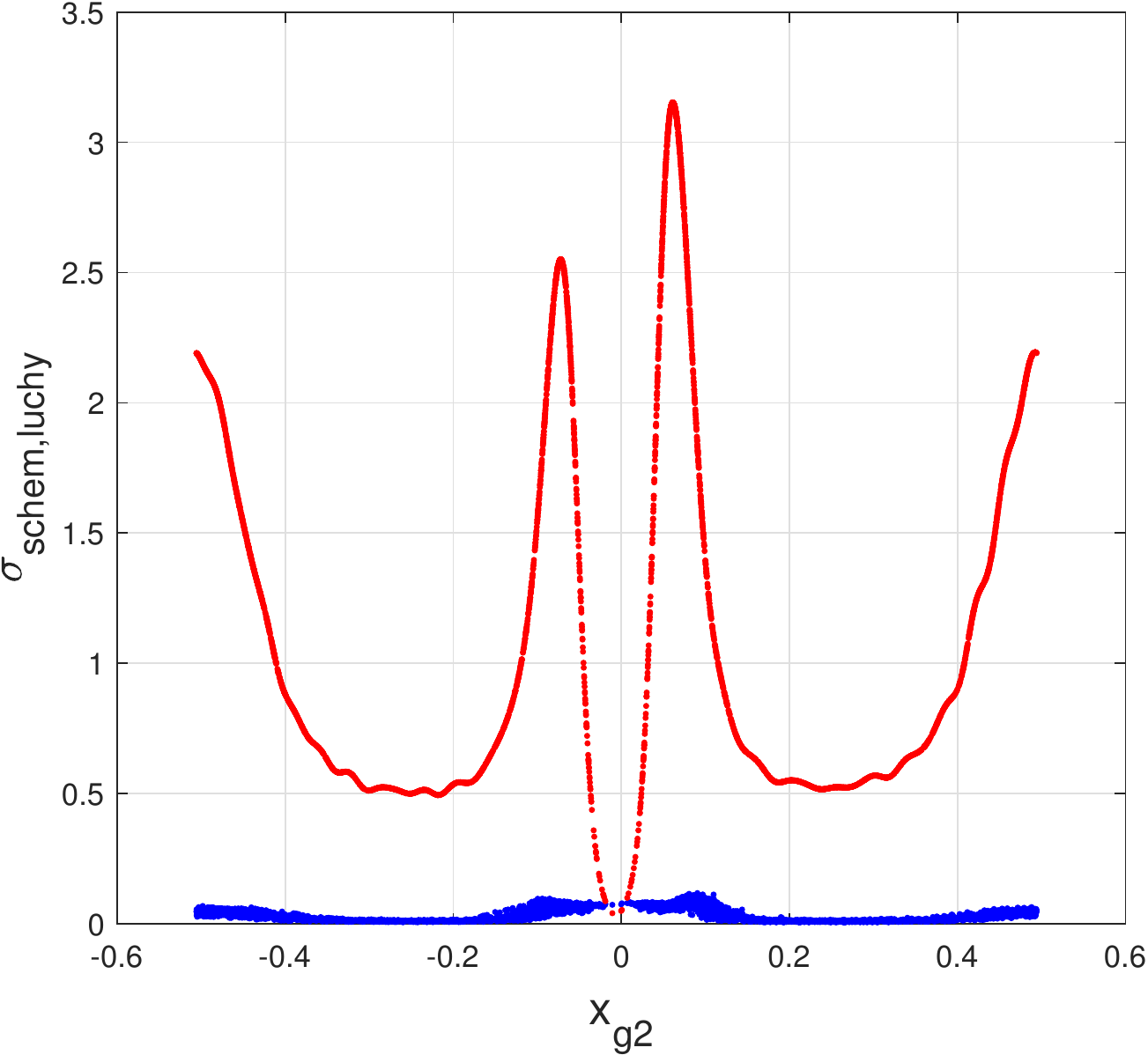}
\includegraphics[scale=0.4]{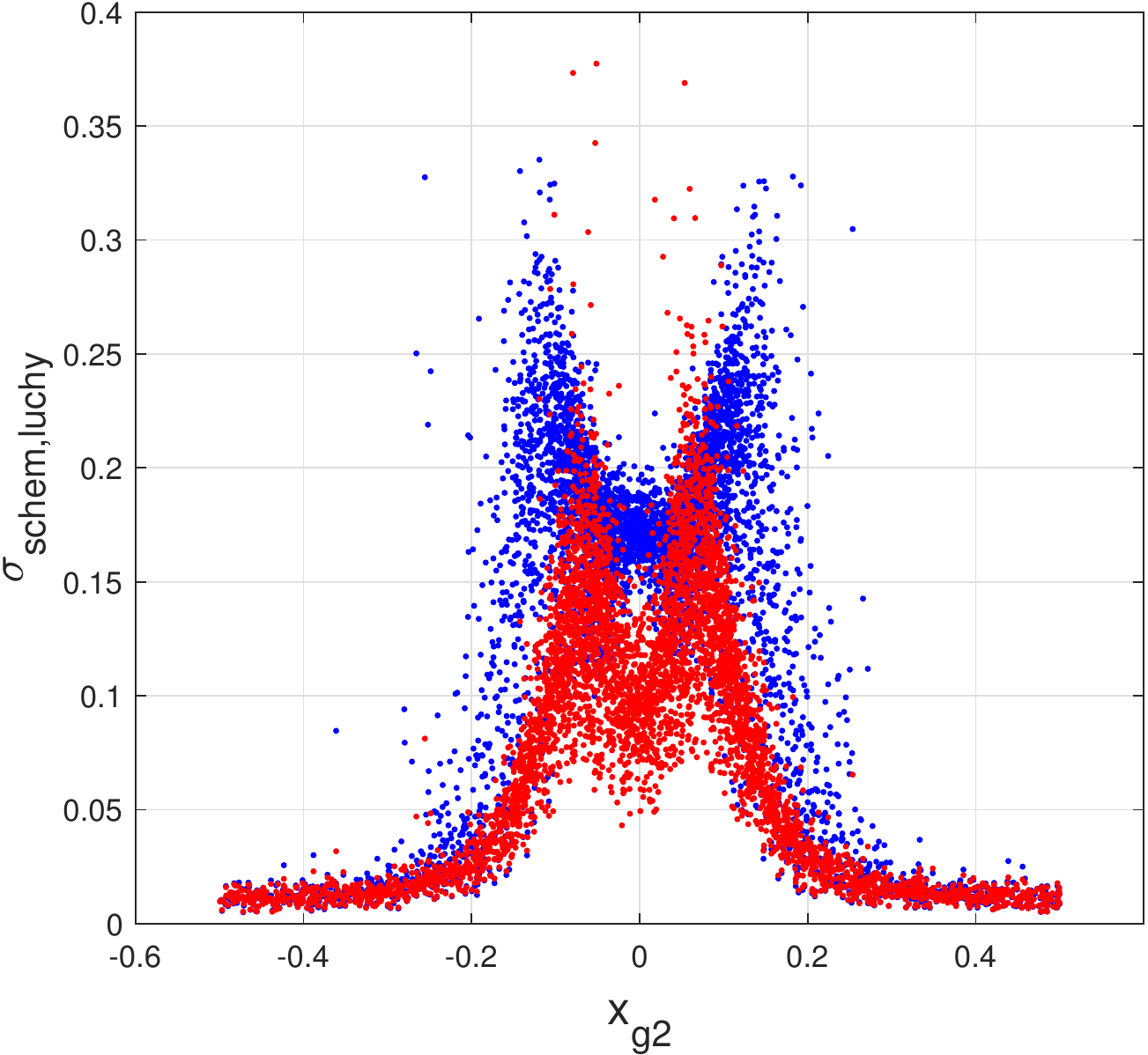}
\includegraphics[scale=0.4]{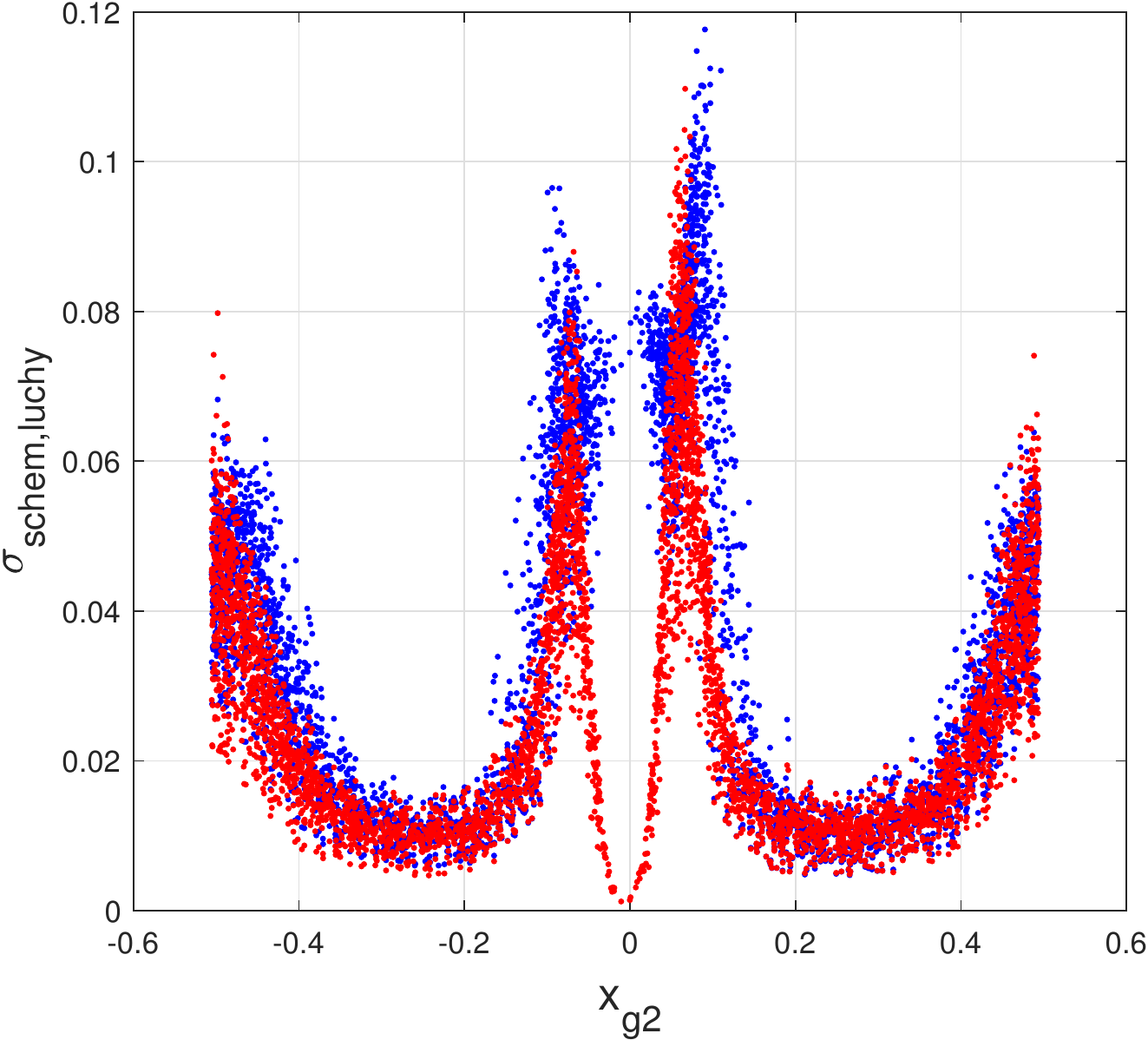}
\caption{ Upper plots:  Blue dots: schematic model, Red dots COG$_2$
histograms. left side plot: normal strip detector.
Right side plot: floating strip detector. Lower Plots:
Blue dots: schematic model, Red dots $\sigma_\eta(x_{g2})$
of equation~\ref{eq:sigma_lucky}. left side plot: normal strip detector.
Right side plot: floating strip detector}
\label{fig:figure_4}
\end{center}
\end{figure}
The upper plots of figure~\ref{fig:figure_4} illustrates
the relations of the effective standard deviations
of~\cite{landi05,landi06} for the two schematic
models and the corresponding amplitudes of the lucky-model.
The lower plots of figure~\ref{fig:figure_4} show the
(surprising) good overlaps of the guessed weights
of equation~\ref{eq:sigma_lucky} for the advanced
lucky-model (hereafter super-lucky model)
with the two schematic models of~\cite{landi05,landi06}.
Also the limitation to an identical set of detectors
is solved by the super-lucky model.
\begin{figure} [h!]
\begin{center}
\includegraphics[scale=0.52]{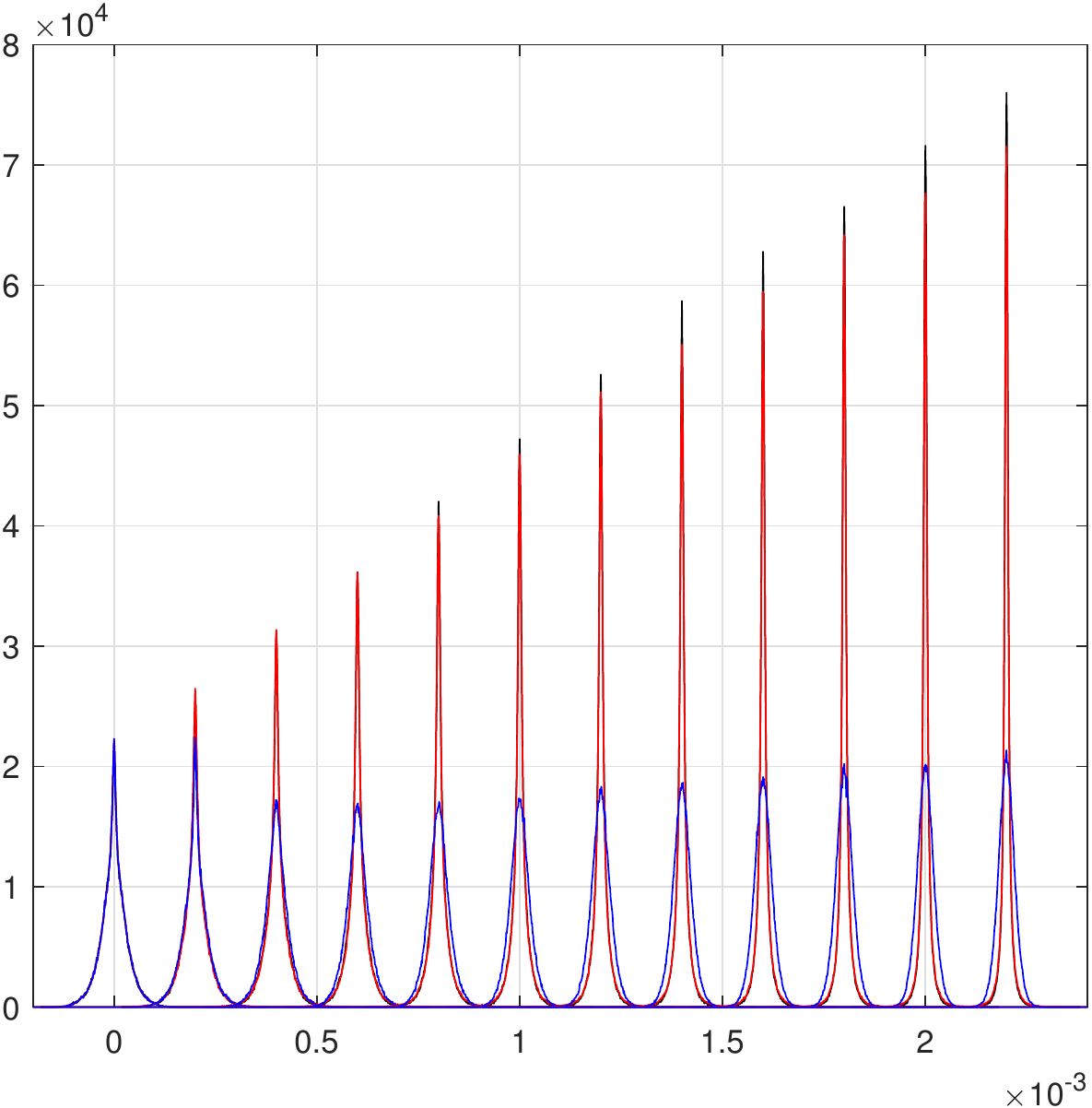}
\includegraphics[scale=0.52]{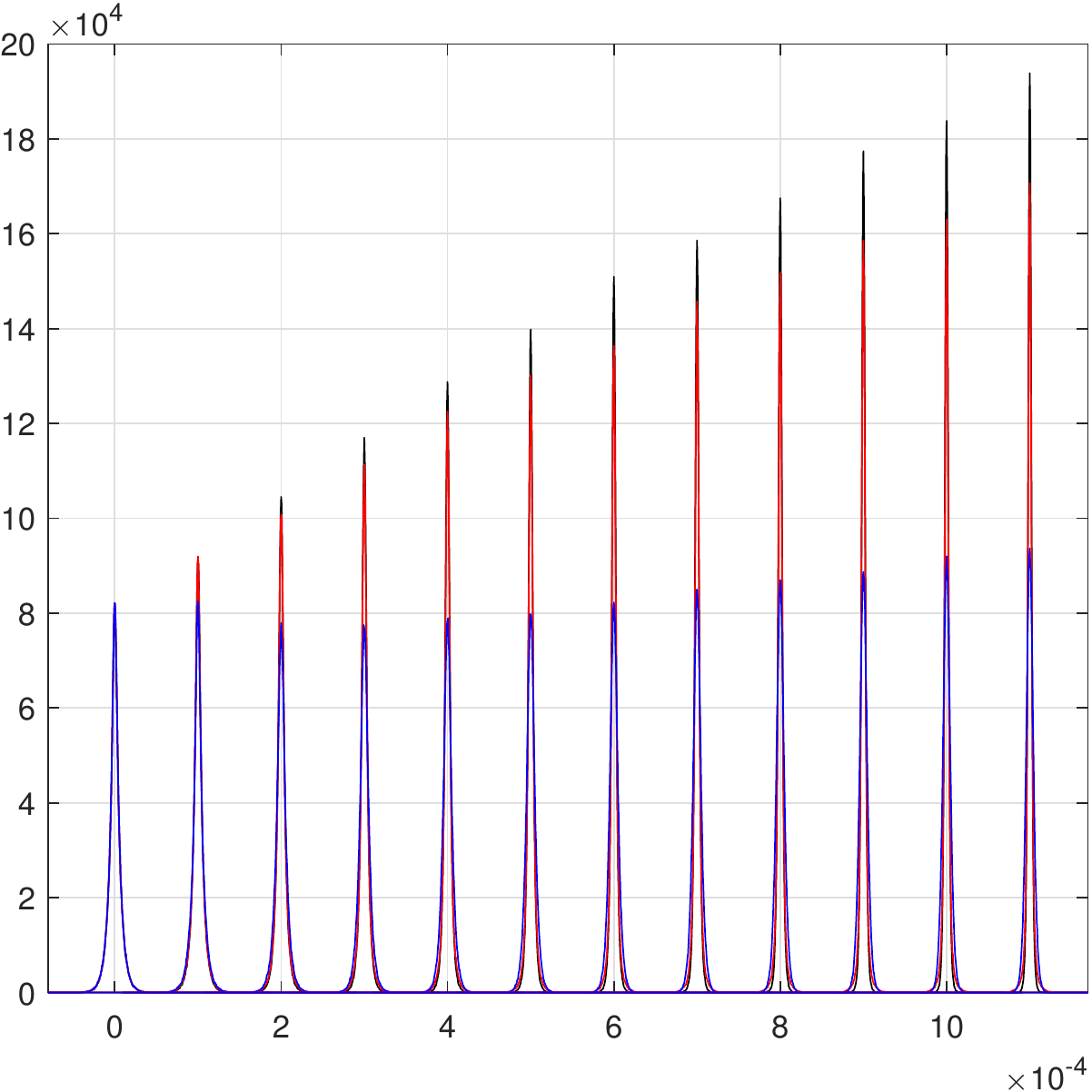}
\caption{ Comparison of the schematic model (black lines),
the super-lucky-model (red lines) and the standard model
(blue lines) for the track direction reconstructions with 2, 3
$\cdots$, 13 detectors layers. Left side plot: normal strip
detectors, right side plot: floating strip detectors}
\label{fig:figure_5}
\end{center}
\end{figure}

Figure~\ref{fig:figure_5}  shows the quality of the
track reconstructions produced by the super-lucky
model that is very near to the schematic model,
substantially better than the simple lucky model
(not reported in figure for a better readability).
Similarly to~\cite{landi15}, figure~\ref{fig:figure_5}
reports the empirical PDFs for the fitted directions of 150,000 straight
tracks at orthogonal incidence on a set of parallel and
equidistant detector layers. To clearly observe
the differences of the parameter distributions, the first
distributions with two detector layers are centered on zero
(as it must be) the other distributions with N=3,
N=4, $\cdots$, up to N=13 are shifted by N-2 identical steps.
We start always from two detector layers as a check
of the methods, all of them must coincide.

An interesting comparison is the relations of the maxima
of the distributions of figure~\ref{fig:figure_5} and the
standard deviations of those distributions. We have to recall
the possible complications with the standard deviation
for systems with PDFs having tails similar the Cauchy-Agnesi PDFs.
As in ref.~\cite{landi09,landi15} we report $1/\sqrt{2\,\pi\,S_d^2}$ where
$S_d$ is the standard deviation of
one of the distributions in figure~\ref{fig:figure_5}. For Gaussian PDF
this ratio coincides with the maximum of the PDF. As
expected by the  (Cauchy-Agnesi) tails of the PDFs,
we observe in these plots large distances from the maxima. The
maxima of the standard least squares are the nearest.
\begin{figure} [h!]
\begin{center}
\includegraphics[scale=0.52]{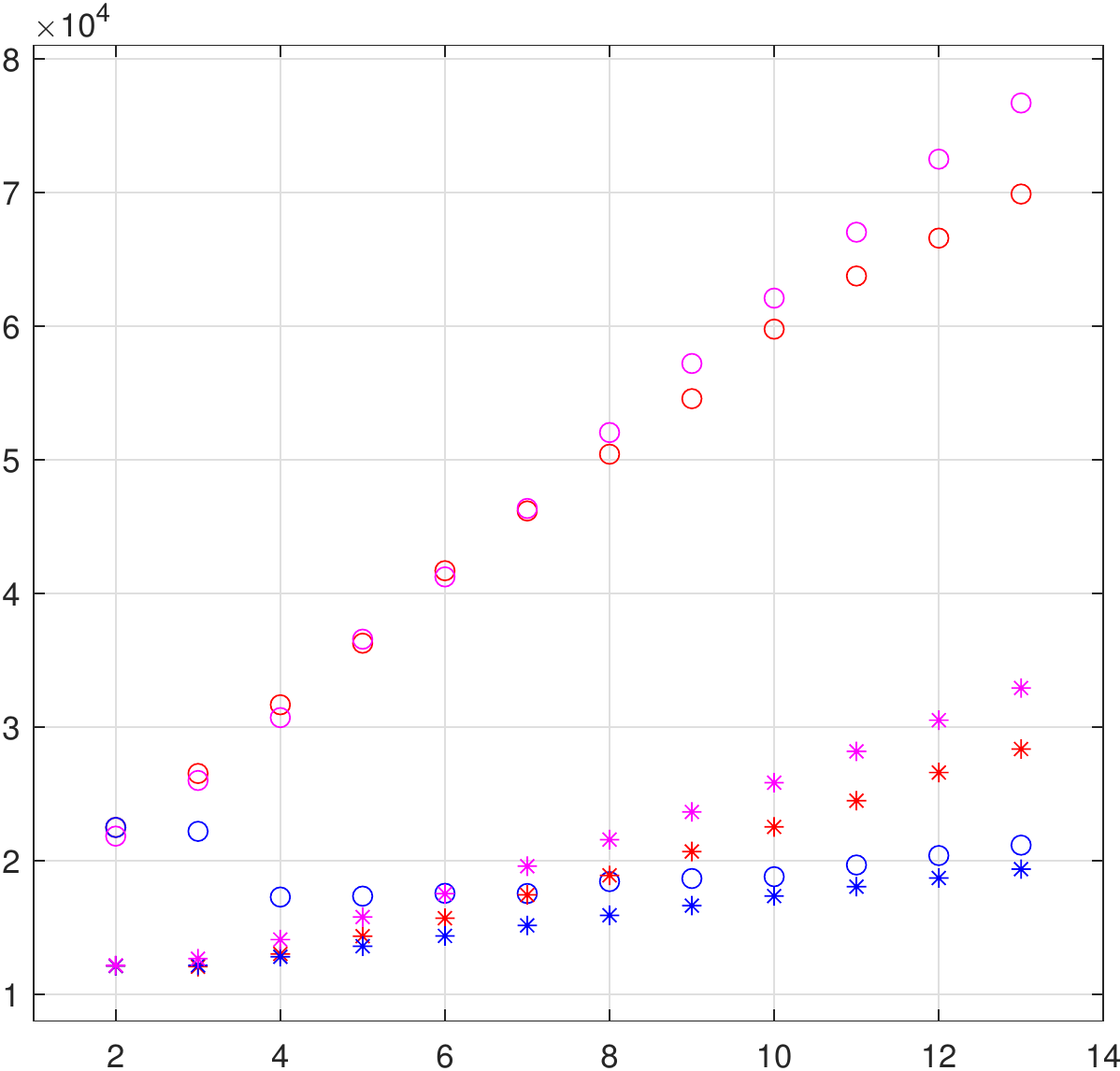}
\includegraphics[scale=0.52]{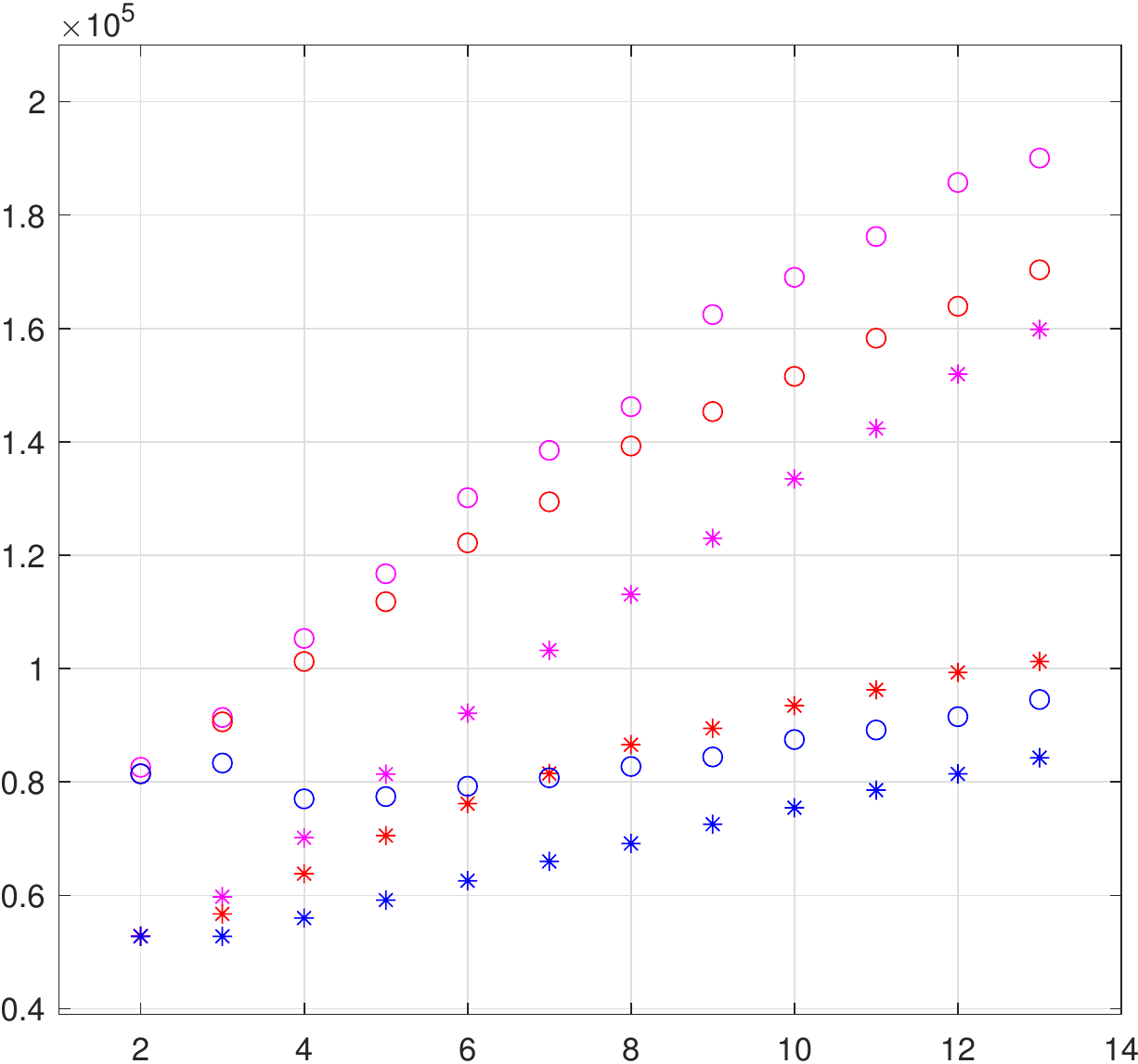}
\caption{ Comparisons of maxima and $1/\sqrt{2\,\pi\,S_d^2}$, $S_d$
the standard deviation of one of the three distributions. The magenta
color refers to the schematic model, the red color for the super-lucky model
and the blue color for the standard fit. The circles indicated the maxima
and the asterisks are for ratio of the standard deviations. The left plot
is for the normal strip case. The right plot is for the floating strips detectors.
}
\label{fig:figure_6}
\end{center}
\end{figure}

\noindent
The results of the lucky model is not reported to avoid
an excessive complication in the plots However, the
simulations with the lucky model for normal strip
detector are reported in ref.~\cite{landi15}. The
results of the lucky model for the floating strip
detector are appreciably lower ($\approx 12\%$) than
the results of the super-lucky model.

\subsection{The super-lucky model for the combination of two very different detector types}

The super-lucky model was studied just for the
application of the lucky model to trackers with
different types of detector layers. Thus, we simulate
a set of trackers starting with two detector layers,
one for each type, and adding alternating a floating
strip detector and a normal detector.
Figures~\ref{fig:figure_5} and~\ref{fig:figure_6}
show the large differences in resolution of these two
types of detectors composing this new system; the
floating strip detectors with
a low noise of four ADC counts and the normal detectors
with a noise of eight ADC counts. The floating strips
add further improvement in resolution for their spreading
the signal in the nearby strips.

\begin{figure} [h!]
\begin{center}
\includegraphics[scale=0.52]{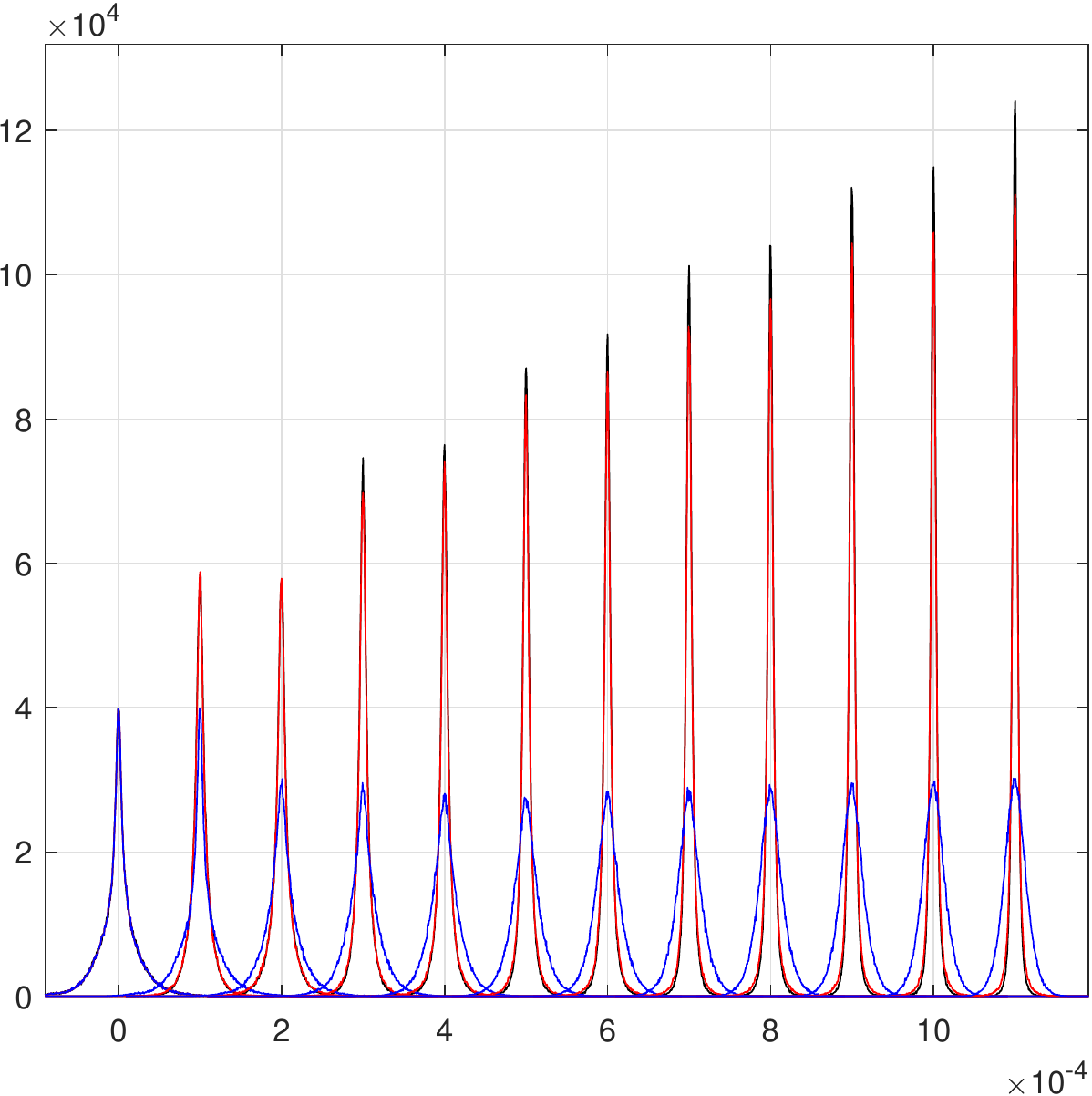}
\includegraphics[scale=0.52]{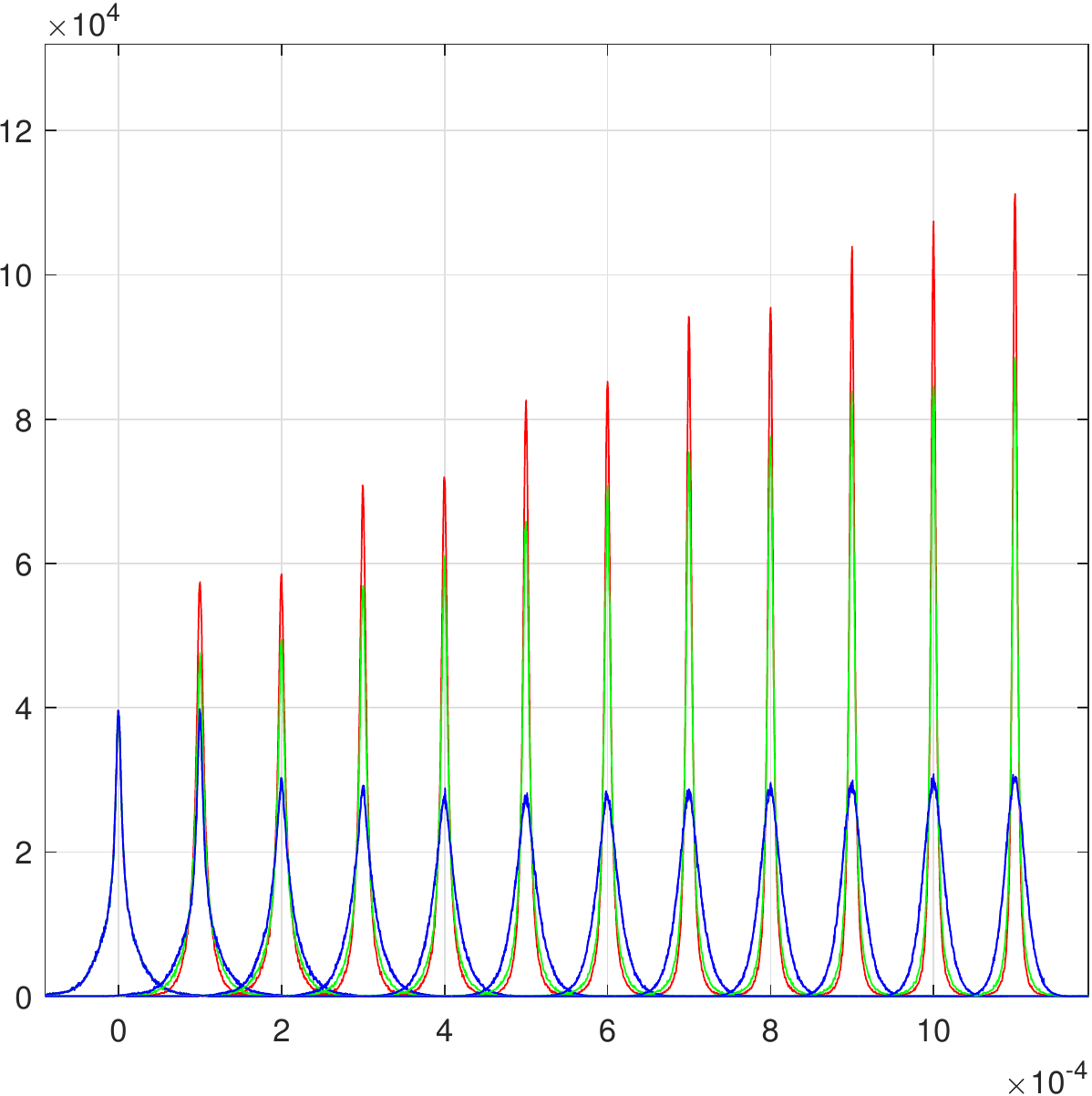}
\caption{ Left plot: Comparison of the schematic model
(black lines), super-lucky-model (red lines)
and the standard model (blue lines) for the track direction
reconstructions half detector layers of floating strip
detector and half with normal detector layers. Left side
plot: always half and half detector types, the green
lines are given by the simpler lucky model, the other
lines are those of the right plot.  }
\label{fig:figure_7}
\end{center}
\end{figure}

\begin{figure} [h!]
\begin{center}
\includegraphics[scale=0.52]{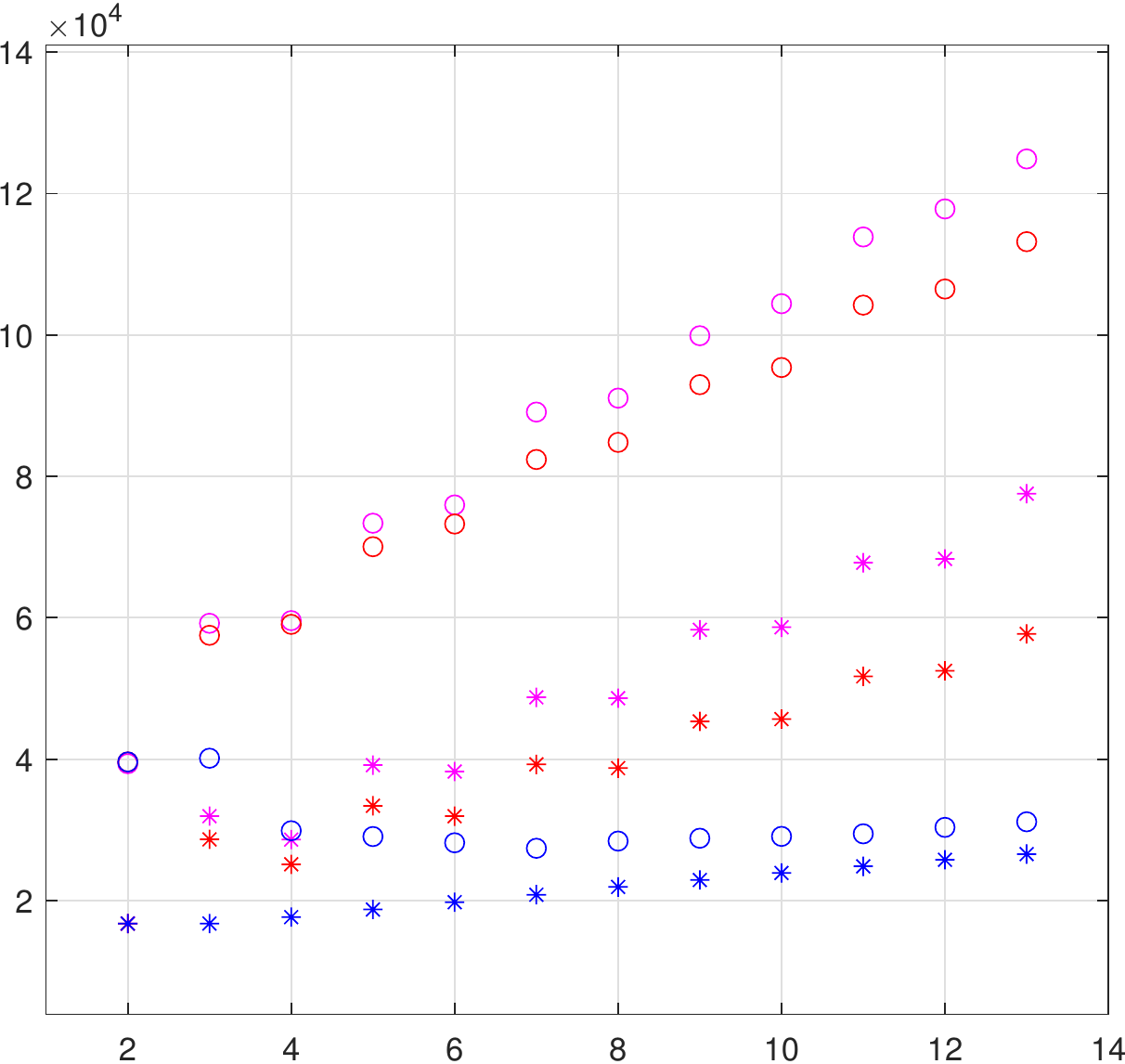}
\includegraphics[scale=0.52]{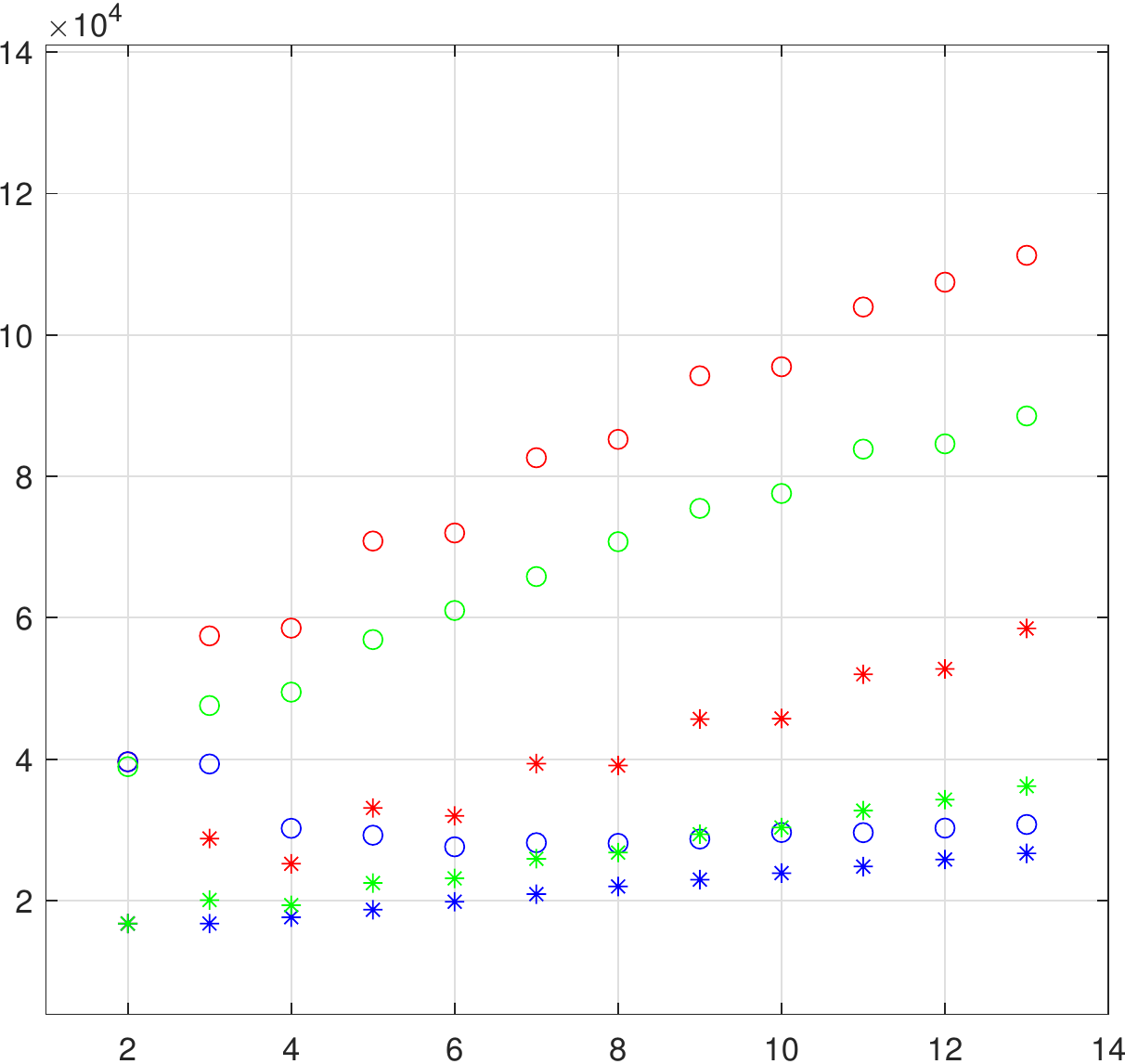}
\caption{ Mixed detector quality. Comparison of maxima and
$1/\sqrt{2\,\pi\,S_d^2}$, $SD$ the standard deviation of one of
the three distributions. The magenta color refers to the
schematic model, the red color for the super-lucky model
and the blue color for the standard fit. The circles
indicated the maxima and the asterisks are for ratio
of the standard deviations. The left plot is for the
normal strip case. The right plot is for the comparison
with the simpler lucky model (green color).}
\label{fig:figure_8}
\end{center}
\end{figure}

The left sides of figures~\ref{fig:figure_7}
and~\ref{fig:figure_8} illustrate
the similarity in resolution of the super-lucky model
and the schematic model for this non-homogeneous
set of trackers.
For comparison,  in the right sides of
figures~\ref{fig:figure_7}
and~\ref{fig:figure_8}, we report also the simulation
with the simpler lucky model. The improvements
of the super-lucky model are evident. Despite of
its inconsistency in this case, the simpler
lucky model shows a substantial increase of
resolution compared with the standard least-square.
The rough pieces of information, contained
in the model, are able to enrich the parameter
distributions of exact values. Instead, the standard
deviations of the parameters of this simpler model
are very near (although better) to those of the standard
least-squares, signaling large tails given by
tracks without good hits. The
super-lucky model has standard deviations appreciably
better than the lucky-model, thus the
added corrections~\ref{eq:sigma_lucky} recover part
of the tails. However, the effects due to the
large tails of the distributions can be
attenuated by the selection of tracks with
two or more good hits.


\noindent
In this set of trackers, the growth of the maxima are
dominated by the floating strip detectors, the addition
of a layer of this type of
detectors shows an evident increase of the maxima of the
distributions. Instead, the addition of a layer of normal
detectors has a negligible effect. The usual linear growth of the
two figures~\ref{fig:figure_6} becomes the step like growth of
figure~\ref{fig:figure_7} and figure~\ref{fig:figure_8}.
Also the standard least-square fit shows a drastic reduction of the
maxima and a general deterioration compared to the simpler analysis
of the parts with good detector types. All the other models
show improvements also for the addition of low quality detectors.
In~\cite{landi09}, the resolutions for these non-Gaussian
distributions are defined by the maxima of the distributions:
The higher the maximum the better the resolution. The method of
fitting a Gaussian in the core of the distributions (as for example
in~\cite{CMS_13,CMS_20}) gives too high values for these
very narrow distributions. The demonstrations of~\cite{landi08}
and~\cite{landi09} proves that the standard least squares model
is never optimum outside the homoscedastic systems.
The results of the previous two suboptimal models enforce the
power of those demonstrations. In fact, any deviation from
homoscedasticity, also with weak correlations with the true
variances for the observations, is able to improve the fit
resolution beyond the results of the standard least squares.

All these results are exclusively obtained with the hit
position given by the $\eta$-algorithm as shown in the
two sides of figure~\ref{fig:figure_3}. The use of the
positions given by the COG$_2$-algorithm (as often done)
suppresses totaly the goodness of these results and the
parameter distributions are lower than those of the standard
least-squares fits (even these substantially lower
than those given by the $\eta$-algorithm).

\subsection{Further discussions and comparisons for the simulations and data}

For an happy coincidence we found a very effective approximation
of the schematic model. When we started to write
equation~\ref{eq:lucky_model}, we have no idea of those further
upgrading. They come out from the environment in which we are
writing the equation. Surprising enough were the overlaps
among the effective standard deviations for excellent hits,
illustrated in the low part of figure~\ref{fig:figure_4}.
Further checks showed also that
the differences of these parameters,
for the same hit, are around $\approx 10\%,\,20\%$
for many thousands of excellent hits.
A very improbable random event, thus a mathematical
explanation has to be found. We have to recall
that the construction of the effective variances
in the schematic model contains an arbitrariness
to be fixed. Equation 11 of~\cite{landi06}
reports a detailed description of this construction.
The probability distributions in the integrals are
similar to those at the right side of figures~\ref{fig:figure_2}
and~\ref{fig:figure_2a}, very different from
Gaussian PDFs for the COG$_2$ around zero. Instead, in
the regions of excellent hits for the floating strip
detectors, the simulations show strong similarity
with Gaussian PDFs for the core of the distributions.
The tails are surely different (Cauchy-Agnesi type)
and had to be cut to avoid divergences. The effective
variances showed a sensible dependency from the cut
position. Our decision about the cut was essentially
"aesthetical", the effective variances had to reproduce
the core with Gaussian PDFs for a small number (four
or five) of excellent hits where that core was
very similar to a Gaussian.
At the same time, the systematic comparison of the
real PDFs with Gaussian PDFs gives indications of
deviations from optimality. In fact, the Gaussian PDFs
are the optimum for the schematic model.
Evidently this selection produces effective standard deviations
very near to those of the super-lucky model
almost perfect for excellent hits. This type of
hits turned out to be far more numerous~\cite{landi05}
than our best expectation.

In any case, a different selection of the cuts can
change this convergence and perhaps the fit quality
for the loss of optimality for those excellent hits.
At the moment of these selections we were in a very
early stage of the work, with no idea of the results.

\begin{figure} [h!]
\begin{center}
\includegraphics[scale=0.52]{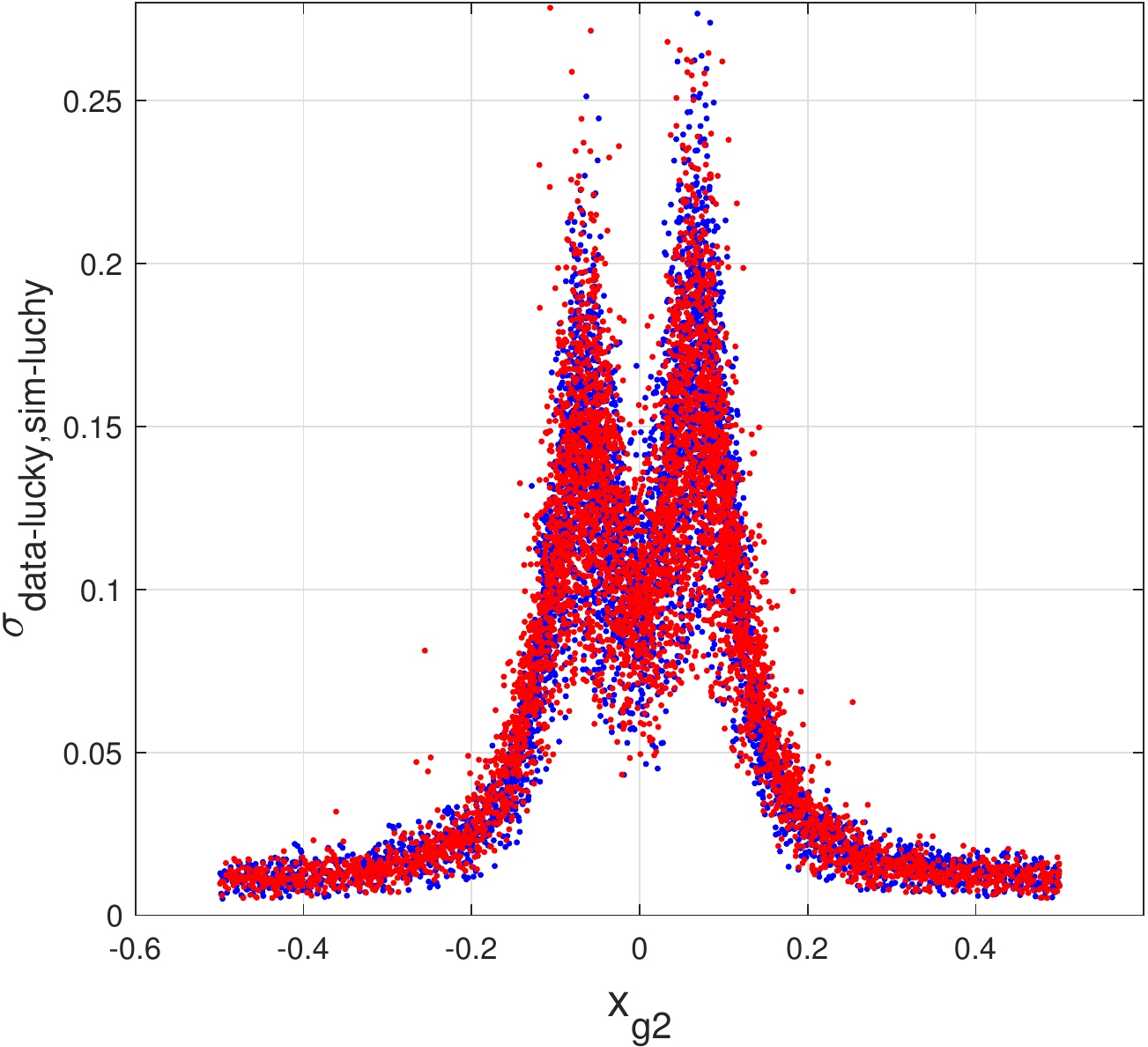}
\includegraphics[scale=0.52]{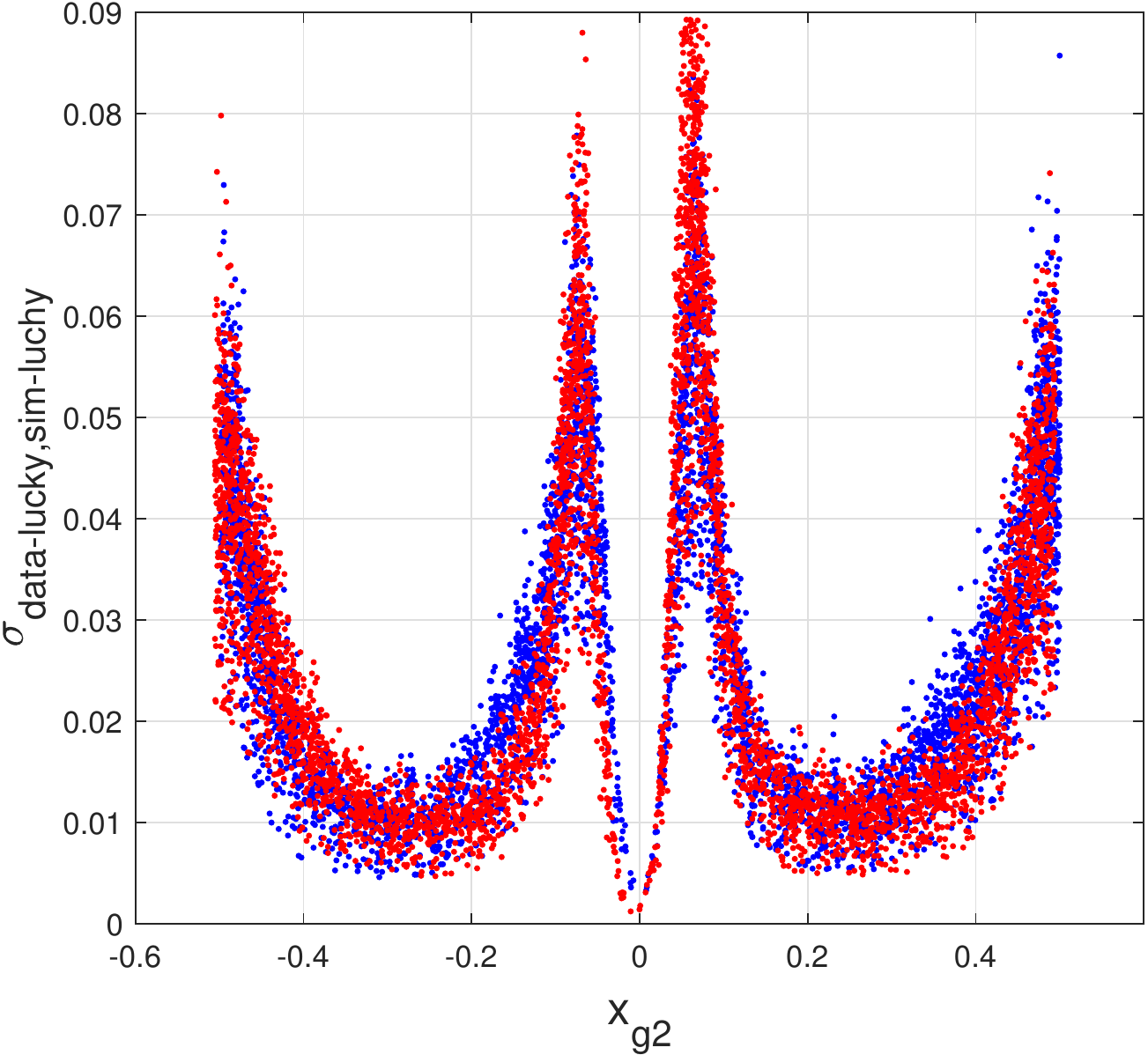}
\caption{Blue dots: $\sigma_\eta(x_{g2})$ of
equation~\ref{eq:sigma_lucky} obtained from the data of a test
beam~\cite{pamela}, Red dots $\sigma_\eta(x_{g2})$
of equation~\ref{eq:sigma_lucky} from simulations.
Left side plot: normal strip detector.
Right side plot: floating strip detector}
\label{fig:figure_10}
\end{center}
\end{figure}

Other interesting comparisons are possible: the $\sigma_\eta(x_{g2})$
of the lucky model from our simulations and the data. The
calculation of these expressions are very fast
(contrary to the schematic model that requires a huge number
of numerical integrations). Figure~\ref{fig:figure_10} shows
these comparisons for the two types of detectors.
The blue dots are now the $\sigma_\eta(x_{g2})$ of the super-lucky model
given by the data from a test-beam~\cite{pamela} and the
red ones are those of the low side of figure~\ref{fig:figure_4}.
The overlaps are excellent, showing the high quality of
our method of simulation~\cite{landi05}.
Although,  the histograms of the simulated signal distributions
on the strips show excellent overlaps with those from data,
this type of comparison was never explored before
for evident reasons. The $\sigma_\eta(x_{g2})$-values
have a non trivial relation with the data
and these strict correlations enforce our system of
extracting/defining~\cite{landi05} the average signal
distribution on the strips.

It is reasonable to expect that
the use of the $\sigma_\eta(x_{g2})$-values, in real track
fitting,  give results comparable to those
of figures~\ref{fig:figure_5} and~\ref{fig:figure_6}
for that detector types
or combinations (figures~\ref{fig:figure_7} and~\ref{fig:figure_8}).

\section{Conclusions}

The analytical expressions for the two strip center of gravity
allow an explanation of the a suboptimal handling of
the heteroscedasticity for
silicon micro-strip detector: the lucky
model. This explanation open the way to an advanced form; the super-lucky model.
This advanced form is able to extend the lucky model to
trackers composed by detectors with very different
properties. The reported simulations show a
substantial increase of the parameter resolution of the
super-lucky model, well beyond the results of the standard least squares
and very near to those of the schematic model.
This method, for the easy availability of its composing
elements, adds negligible complications to the fitting work
with substantial increases of resolution.



\end{document}